\documentclass{aastex6}
\usepackage{xcolor}
\fullcollaborationName{The TDSS Collaboration}
\begin{document}

\title{The Time-Domain Spectroscopic Survey: TARGET SELECTION FOR REPEAT SPECTROSCOPY}

\author{Chelsea L. MacLeod\altaffilmark{1}}
\author{Paul J. Green\altaffilmark{1}}
\author{Scott F. Anderson\altaffilmark{2}}
\author{Michael Eracleous\altaffilmark{3,4}}
\author{John J. Ruan\altaffilmark{2}}
\author{Jessie Runnoe\altaffilmark{5}}
\author{William Nielsen Brandt\altaffilmark{3,4,6}}
\author{Carles Badenes\altaffilmark{7}}
\author{Jenny Greene\altaffilmark{8}}
\author{Eric Morganson\altaffilmark{9,10}}
\author{Sarah J. Schmidt\altaffilmark{11}}
\author{Axel Schwope\altaffilmark{11}}
\author{Yue Shen\altaffilmark{10,9,12}}
\author{Rachael Amaro\altaffilmark{1,10}}
\author{Amy Lebleu\altaffilmark{1,13}}
\author{Nurten Filiz Ak\altaffilmark{14}}
\author{Catherine J. Grier\altaffilmark{3,4}}
\author{Daniel Hoover\altaffilmark{2}}
\author{Sean M. McGraw\altaffilmark{3,4}}
\author{Kyle Dawson\altaffilmark{15}}
\author{Patrick B. Hall\altaffilmark{16}}
\author{Suzanne L. Hawley\altaffilmark{2}}
\author{Vivek Mariappan\altaffilmark{15}}
\author{Adam D. Myers\altaffilmark{17}}
\author{Isabelle P\^aris\altaffilmark{18}}
\author{Donald P. Schneider\altaffilmark{3,4}}
\author{Keivan G. Stassun\altaffilmark{19,20}}
\author{Matthew A. Bershady\altaffilmark{21}}
\author{Michael R. Blanton\altaffilmark{22}}
\author{Hee-Jong Seo\altaffilmark{23}}
\author{Jeremy Tinker\altaffilmark{22}}
\author{J. G. Fern\'andez-Trincado\altaffilmark{24,25}}
\author{Kenneth Chambers\altaffilmark{26}}
\author{Nick Kaiser\altaffilmark{26}}
\author{R.-P. Kudritzki\altaffilmark{26}}
\author{Eugene Magnier\altaffilmark{26}}
\author{Nigel Metcalfe\altaffilmark{27}}
\author{Chris Z. Waters\altaffilmark{26}}

\altaffiltext{1}{Harvard Smithsonian Center for Astrophysics, 60 Garden St, Cambridge, MA 02138, USA}
\altaffiltext{2}{Department of Astronomy, University of Washington, Box 351580, Seattle, WA 98195, USA}
\altaffiltext{3}{Department of Astronomy \& Astrophysics, 525 Davey
  Laboratory, The Pennsylvania State University, University Park, PA
  16802, USA} 
\altaffiltext{4}{Institute for Gravitation and the Cosmos, The
  Pennsylvania State University, University Park, PA 16802, USA} 
\altaffiltext{5}{Department of Astronomy, University of Michigan, 1085 S. University Avenue, Ann Arbor, MI 48109, USA}
\altaffiltext{6}{Department of Physics, The Pennsylvania State University, University Park, PA 16802, USA}
\altaffiltext{7}{Department of Physics and Astronomy and Pittsburgh
  Particle Physics, Astrophysics and Cosmology Center (PITT PACC),
  University of Pittsburgh, 3941 O'Hara St, Pittsburgh, PA 15260, USA}
\altaffiltext{8}{Department of Astrophysical Sciences, Princeton
  University, Princeton, NJ 08544, USA}
\altaffiltext{9}{National Center for Supercomputing Applications,
  University of Illinois at Urbana-Champaign,  1205 W. Clark St., Urbana, IL 61801, USA} 
\altaffiltext{10}{Department of Astronomy, University of Illinois at
  Urbana-Champaign, 1002 W. Green Street, Urbana, IL 61801, USA}

\altaffiltext{11}{Leibniz-Institut f\"{u}r Astrophysik (AIP), An der Sternwarte 16, 14482, Potsdam, Germany}
\altaffiltext{12}{Alfred P. Sloan Research Fellow}
\altaffiltext{13}{Department of Physics \& Astronomy, Louisiana State
  Univiersity, 202 Nicholson Hall, Baton Rouge, LA 70803, USA}
\altaffiltext{14}{Faculty of Sciences, Department of Astronomy and Space Sciences, and Astronomy and Space Sciences Observatory and Research Center, Erciyes University, 38039 Kayseri, Turkey}
\altaffiltext{15}{Department of Physics \& Astronomy, University of
  Utah, Salt Lake City, UT 84112, USA}
\altaffiltext{16}{Department of Physics \& Astronomy, York University,
  4700 Keele Street, Toronto, Ontario M3J 1P3, CAN}
\altaffiltext{17}{Department of Physics and Astronomy, University of Wyoming, Laramie, WY 82071, USA}
\altaffiltext{18}{INAF - Osservatorio Astronomico di Trieste, Via G. B. Tiepolo 11, I-34131 Trieste, Italy}
\altaffiltext{19}{Vanderbilt University, Department of Physics \& Astronomy, 6301 Stevenson Center Ln., Nashville, TN 37235, USA}
\altaffiltext{120}{Fisk University, Department of Physics, 1000 17th Ave. N.,Nashville, TN 37235, USA}
\altaffiltext{21}{Department of Astronomy, University of Wisconsin-Madison, 475 N. Charter St., Madison, WI 53706, USA}
\altaffiltext{22}{Center for Cosmology and Particle Physics, Department of Physics, New York University, 4 Washington Place, New York, NY 10003, USA}
\altaffiltext{23}{Department of Physics and Astronomy, Ohio University, 251B Clippinger Labs, Athens, OH 45701, USA}
\altaffiltext{24}{Departamento de Astronom\'ia, Universidad de Concepci\'on, Casilla 160-C, Concepci\'on, Chile} 
\altaffiltext{25}{Institut Utinam, CNRS UMR6213, Univ. Bourgogne Franche-Comt\'e, OSU THETA , Observatoire de Besan\c{c}on, BP 1615, 25010 Besan\c{c}on Cedex, France}
\altaffiltext{26}{Institute for Astronomy, University of Hawaii at Manoa, Honolulu, HI 96822, USA}
\altaffiltext{27}{Department of Physics, University of Durham Science Laboratories, South Road Durham DH1 3LE, UK}

\email{cmacleod@cfa.harvard.edu; pgreen@cfa.harvard.edu}

\begin{abstract}
As astronomers increasingly exploit the information available in the
time domain, spectroscopic variability in particular opens broad new
channels of investigation. Here we describe the selection algorithms
for all targets intended for repeat spectroscopy in the Time Domain
Spectroscopic Survey (TDSS), part of the extended Baryon Oscillation
Spectroscopic Survey within the Sloan Digital Sky Survey-IV. Also
discussed are the scientific rationale and technical constraints leading
to these target selections. 
The TDSS includes a large ``Repeat Quasar Spectroscopy'' (RQS)
program delivering $\sim$13,000 repeat spectra of confirmed SDSS quasars,
and several smaller ``Few-Epoch Spectroscopy'' (FES) programs
targeting specific classes of quasars as well as stars.  
The RQS program aims to provide a large and diverse quasar data set
for studying variations in quasar spectra on timescales of years, a
comparison sample for the FES quasar programs, and opportunity for
discovering rare, serendipitous events.  
The FES programs cover a wide variety of phenomena in both quasars and
stars.  Quasar FES programs target broad absorption line quasars, high
signal-to-noise ratio normal broad line quasars, quasars
with double-peaked or very asymmetric broad emission line profiles,
binary supermassive black hole candidates, and the most photometrically
variable quasars.  Strongly variable stars are also targeted for
repeat spectroscopy, encompassing many types of eclipsing binary
systems, and classical pulsators like RR Lyrae.  Other stellar FES
programs allow  spectroscopic variability studies of active ultracool
dwarf stars, dwarf carbon stars, and white dwarf/M dwarf spectroscopic
binaries. We present example TDSS spectra and describe anticipated
sample sizes and results.
\end{abstract}

\section{Introduction} \label{sec:intro}
With the massive photometric datasets expected from the next
generation of time-domain imaging surveys,  classification 
algorithms will become increasingly dependent on  our understanding
of cosmic variables. Recently, the Sloan Digital Sky Survey (SDSS)-IV
 \citep{blanton17} extended Baryon Acoustic Oscillation Sky Survey
 \citep[eBOSS;][]{dawson16}  has enabled
spectroscopy of celestial variables through  the  Time 
Domain Spectroscopic Survey (TDSS). 
TDSS has been operational since August 2014 and had
obtained 47,000 spectra of stars and quasars as of July 2016. The
target selection for the main TDSS single-epoch-spectroscopy (SES)
program, in which optical point sources (unconfirmed quasars
and stars) are targeted based on variability for a first 
epoch of spectroscopy, prioritized to achieve a typical surface
density on the sky of $\sim$10~deg$^{-2}$,  is described in \cite{mor15}. 
Initial results from a pilot SES survey
during SDSS-III \citep{eisenstein11} are presented in \citet{rua16}.

Aside from the discovery and classification of the variable sky, the
\emph{spectroscopic variability} of some classes of objects is of
considerable interest.  For example, time-domain spectroscopy of broad
absorption line quasars was included in SDSS-III
\citep[e.g.,][]{ak12,ak13,ak14}.  
Rather than be satisfied with extant SDSS
spectroscopy for heterogeneously targeted objects,
TDSS intentionally seeks repeat spectroscopic observations for subsets of known
stars and quasars that are interesting astrophysically via
several ``few-epoch spectroscopy'' (FES) subprograms. 
Keeping within the tight overall TDSS total
 fiber-budget of $\sim10$~deg$^{-2}$, each distinct FES subclass is
 approximately aimed to include of order $10^{2-3}$ objects per FES
 subclass, i.e., the minimum needed for various reasons to achieve
 better than $\sim10\%$ statistics per subclass. With the actual FES
 subclasses implemented, this then means that of order $\sim10\%$ of
 the total TDSS fiber-budget is allotted to FES, yielding an average target 
 density of 1~deg$^{-2}$.

Recently, an eBOSS emission line galaxy (ELG) survey spanning 300
plates began observations in the Fall of 2016
\citep{raichoor17}, covering some areas of the sky
previously observed in SDSS-IV.
 For the TDSS fiber allotment of $\sim10$~deg$^{-2}$ on these plates, we
considered three main options for a targeting strategy: 
($i$) probing deeper into the  photometrically variable  density,
confirming new SES variables with enhanced completeness but  reduced purity,
($ii$) re-observing previous SES targets,
therefore obtaining repeat spectroscopy for variable quasars and stars in general, or
($iii$) shifting to a greater quasar emphasis, targeting more quasars
previously observed in those fields.
Since the ELG survey overlaps existing SDSS-IV fields
as well as regions not yet covered by SDSS-IV, 
we adopted a new target selection for these plates, and   
chose to acquire repeat spectra of quasars already known in the field (option $iii$); this choice also serves as a pilot for a potentially larger
program in future all-sky spectroscopic surveys. 
In this paper, we describe all the non-SES TDSS sub-programs
that select objects for multiple spectroscopic observations, 
including several smaller FES programs
covering the  full SDSS sky area, and this pilot program of repeat quasar
spectroscopy (RQS) on the ($\approx$1200 deg$^2$) area encompassed by the ELG plates.

Astrophysically, the main contributors to the variable sky, and thus
the TDSS repeat spectroscopy sub-programs, are quasars and variable
stars.   The hallmarks of quasar spectra include the
power-law continuum from a thermally emitting
accretion disk, broad emission lines (BELs) from the 
broad line region (BLR) that are photoionized by a higher-energy UV
continuum \citep{pet93}, and narrow emission lines 
\citep[see review by][]{ost86}.
The Balmer lines  (H$\alpha$, H$\beta$, etc.) 
have historically been extremely useful for inferring information
about the physical structure and dynamics of the BLR, and are directly
related to the number of ionizing photons from the continuum source \citep{kor04}.
In rare cases, these lines can be double peaked when
feeble winds allow a low optical depth sightline to the outermost part
of the BLR disk \citep{eracleous09}.  The Balmer lines also form the
basis of our 
``Type I'' vs.\ ``Type II'' classification scheme, and repeat
spectroscopy has been useful in the past to identify contaminants
(e.g., with weak, broad Balmer emission lines) in 
samples of Type II quasars \citep{barth14}.  
The formation of other BELs is thought to involve more
complicated processes \citep{waters16}. For example, the \ion{Mg}{2} BEL is subject to 
collisional de-excitation and can show different emission properties than the Balmer
lines \citep{roig14,cac15,sun15}. 
The UV lines (e.g., \ion{C}{4}, \ion{N}{5}) are known to trace winds \citep{richards11}.
Broad absorption lines (BALs), found in 10-25\% of quasars
\citep[e.g.,][]{wey91,trump06,gib09}, are believed to be formed in 
a wind that is launched from the accretion disk at 10--100 light days
from the supermassive black hole \citep[e.g.,][]{mur95,pro00}.  A larger fraction of quasar
spectra show  narrow absorption lines \citep[NALs;
e.g.,][]{lundgren09}, and some have absorption lines of intermediate
widths (mini-BALs).\footnote{Operationally, BALs have widths $>$ 2000 km/s, NALs have widths $<$ 500 km/s, 
and   mini-BALs   are   in-between.}   The existence of BAL variability that has been
systematically studied since SDSS-III \citep{ak12,ak13,ak14} has been
known for almost three decades \citep[see review by][]{tur88}.  

The menagerie of stellar variable classes encompasses causes both
intrinsic (e.g., pulsators like RR Lyrae, Cepheids, and long-period
variables and eruptive stars like CVs, novae, and symbiotics) and
observer-dependent (e.g., eclipsing and/or spectroscopic binaries and rotation)
variables. Photometrically, a few percent of all stars are
considered variable, with the exact number depending on   
the filter bands used, the cadence of observations, and the limiting
magnitude of the sample.
 With dense photometric monitoring,  
a few percent of those will show periodic variability (arising from
radial pulsations, rotation or orbital motion). The 
remaining variable stars exhibit eruptive or irregular variability, the
latter class including flaring stars across the main sequence
\citep[e.g.,][]{Davenport2016}.
The periodic stellar variables can be classified with
reasonable efficiency based on their period, amplitude, and lightcurve
shape \citep[e.g.,][]{palaversa13,drake14,vanderplas15}.   Spectroscopy provides
further important physical characteristics such as gravity,
temperature, and radial velocity.  However, single-epoch spectroscopy
captures but a single phase, and may provide little insight into the
physical reason for the observed variability.   Multi-epoch spectra,
in contrast, can provide key information about e.g., radial velocity
variations and orbital properties of binaries, or emission line
variability related to chromospheric activity, irradiation, or
accretion.   Spectroscopic variability surveys in broad stellar
samples are virtually nonexistant to date, but can broadly
characterize stellar variability in physical detail, and also focus on
specific classes of stars that are known or suspected to be variable. 

We describe the TDSS input data sets used for choosing our spectroscopic
targets in \S\ref{sec:data}. In the following sections, we present
the selection algorithms used by various 
subprograms of TDSS to target objects for repeat spectroscopy, with
the FES programs described in \S\ref{sec:fes} and
the RQS program in \S\ref{sec:rqs}.  We
summarize these programs in \S\ref{sec:summary}.

\section{Input Data Sets}
\label{sec:data}

Spectra from the first two phases of the SDSS \citep[][also referred to
as SDSS-I/II]{york00} form the basis of most samples targeted here. 
The SDSS legacy survey includes observations up
through Data Release (DR) 7 \citep{DR7}.  The SDSS-III survey
 continued to extend the imaging and spectroscopic
sky coverage of the SDSS surveys, culminating with DR12 \citep{alam15}. 
The SDSS-IV project eBOSS began in July 2014, marking the
formal start of TDSS observations. The most recent data release is DR13  \citep{DR13}.

We use imaging data from the SDSS-I/II/III and Pan-STARRS-1
\citep[PS1;][]{kai02} 3$\pi$ surveys. 
SDSS started its imaging campaign in 2000 and concluded in 2007, having covered
11,663 deg$^2$.  SDSS-III added $\sim$3000 deg$^2$ of new imaging area in 2008. 
PS1 imaging commenced in 2009 and provided light curves for all SDSS
sources through 2013.  Hence, the addition of the PS1 photometry to
the SDSS photometry increases the span of light curves from $\approx$8 to $\approx$14 years.
More details on each survey and the photometric variability measures
used in (\S\ref{sec:fes:HYP}) and (\S\ref{sec:rqs}) are described in this
section.

\subsection{SDSS Imaging}
    The SDSS uses the imaging data gathered by a dedicated
    2.5m wide-field telescope \citep{gunn06}, which collected light from a
    camera with 30 2k$\times$2k CCDs \citep{gunn98} over five broad bands
    - {\it ugriz} \citep{fukugita96,doi10} - in order to image 14,555 unique
    deg$^{2}$ of the sky. This area includes \hbox{7,500 deg$^{2}$} in the
    North Galactic Cap (NGC) and \hbox{3,100} deg$^{2}$ in the South
    Galactic Cap (SGC). 
    The Eighth Data Release \citep[DR8;][]{DR8} provides the full imaging data set and updated
    photometric calibrations. This catalog provides the magnitudes and
    astrometry used in constructing our target samples; all coordinates
    hereafter are J2000.

    The Stripe 82 region of SDSS (S82; $-60^{\circ} < \alpha <50^{\circ}$
    and $|\delta|<1.27^{\circ}$) covers $\sim$300~deg$^2$ and has been observed
    $\sim$60 times on average to search for transient and variable objects \citep{frieman08,DR7}.  These
    multi-epoch data probe time scales ranging from 3 hours to 8 years and
    provide well-sampled 5-band light curves. The S82
    variable and standard star catalogs \citep{ivezic07} are used to
    train the TDSS SES selection in
    \cite{mor15}, and therefore the hypervariables selection
    (\S\ref{sec:fes:HYP}). The S82 data are used for variability
    selection in the RQS program (\S\ref{sec:rqs}).

\subsection{Pan-STARRS1 3$\pi$ Survey}
PS1 utilizes a 1.8m telescope equipped with a 
1.4-gigapixel camera.  Over the course of 3.5 years of the 3$\pi$ survey, up to four
exposures per year in 5 bands, $g_{\rm PS1}, r_{\rm PS1}, i_{\rm PS1},
z_{\rm PS1}, y_{\rm PS1}$ have been taken across the entire $\delta >
-30^{\circ}$ sky \citep[for full details, see][]{Tonry12,met13}.  
Each nightly observation consists of a pair of exposures separated by 15~min  
to search for moving objects. For each exposure, the PS1 3$\pi$ 
survey has a typical 5$\sigma$ depth of 22.0 in the $g$-band \citep{ins13}.
The instrumentation, photometric system, and  the PS1 surveys  are described in
\citet{Kaiser10}, \citet{Stubbs10}, and \citet{Magnier13},
respectively. A high quality subset of PS1 data was first released in 
Processing Version (PV) 1, and PV2 added data through a later date in the observing, 
as well as some previously missed earlier observations via better
analysis failure handling \citep[for an overall description of
the PS1 database, see][]{flewelling16}. 

\subsection{Variability Measures}
\label{sec:data:var}
Whereas variability selection formed the basis for the first epoch
(SES) target selection described in \cite{mor15}, multiepoch imaging
data were also used to select known stars and quasars for the  
extremely variable (or ``hypervariable''; \S\ref{sec:fes:HYP}) and RQS
quasar (\S\ref{sec:rqs}) samples. The SES selection uses a three-dimensional
parameter space (magnitude, PS1-only variability, and SDSS--PS1 difference), 
designed to achieve a high-purity variable sample at a typical surface
density on the sky of $\sim$10~deg$^{-2}$.
As part of the SES  selection, hypervariables were selected 
based on a single variability metric $V$ that parameterizes variability in a
two-dimensional space of  two variability terms $S_1$ and $S_2$:
\begin{eqnarray}
V &=& \left(\rm{median}(|\rm{mag}_{\rm{PS1}}-\rm{mag}_{\rm{SDSS}}|)^2+4\ \rm{median}(Var_{\rm{PS1}})^2\right)^{1/2} = \left(S_1^2+4 S_2^2\right)^{1/2},\label{eq:V}
\end{eqnarray}
\citep[Equation 11 of][]{mor15}. Qualitatively, $S_1$ is the PS1--SDSS
difference and represents long term (multi-year) variability, and
$S_2$ is the PS1-only variability characterizing short term (days to a
few years) variability.  
 This non-standard variability measure was intended to combine short term
 variability and long term variability measures into one quantity.
 The training sets used to derive this quantity, as well as the
 detailed statistics of this selection can be found in \cite{mor15}.
Hypervariables with $V>2$~mag were
prioritized during the SES selection, whereas for the FES targets, a lower
$V$ threshold is used (see \S\ref{sec:fes:HYP}).  
As for the SES selection, an updated version of the ``ubercalibrated'' PS1
data \citep{schlafly12}, which include PV1 data up through July 2013, 
was used to select FES hypervariables. For
detailed definitions, see \citet{mor15}.

\begin{figure}
\plotone{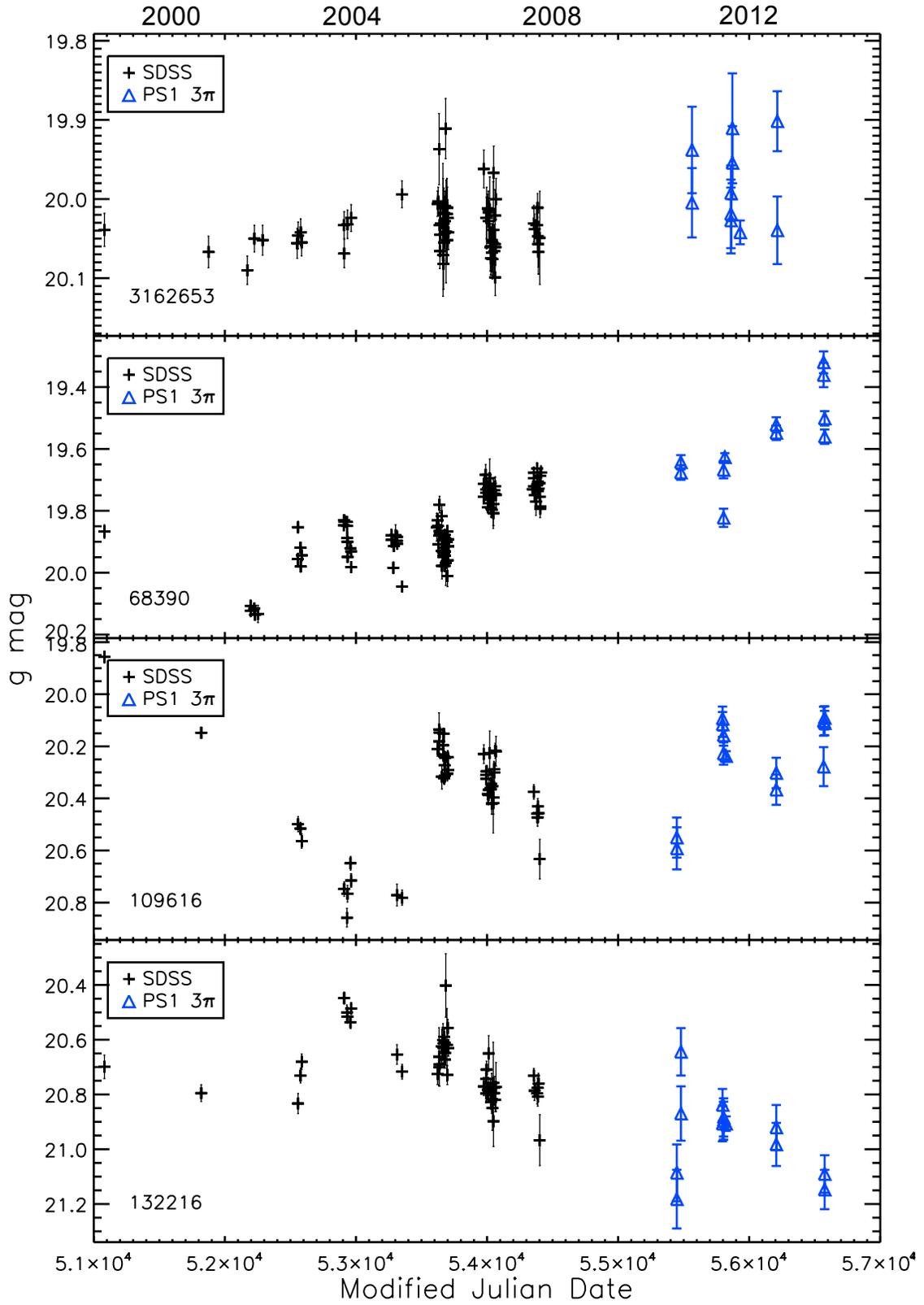}
\caption{Example light curves of known quasars in Stripe 82. The first panel shows an
  example with relatively insignificant
  variability ($\chi^2_{pdf}<3$). The quasars in the remaining panels show
  highly significant variability ($\chi^2_{pdf}>30$) and are thus selected as 
  RQS targets (see \S\ref{sec:rqs}). \label{fig:lceg}}
\end{figure}

The RQS program targets known SDSS quasars based on median SDSS magnitude and highly significant
variability (see \S\ref{sec:rqs}).  Here, the 
variability selection is based on the reduced $\chi^2$ of the light curve:
\begin{eqnarray}
\chi^2_{pdf} = \frac{1}{n-1}  \displaystyle\sum_{i=1}^{n}{ \left(\frac{ m_i - \mu }{ \sigma_i}\right)^2},
\end{eqnarray}
where $n$ is the number of data points in a given filter, $m_i$, ..., $m_n$ are the
individual magnitudes, $\sigma_i$ is the error associated with $m_i$,
and $\mu$ is the mean magnitude.
The reduced $\chi^2_{pdf}$ in both $g$ and $r$ bands is used to
define the variability-selected subsamples  (see \S\ref{sec:rqs}
  for detailed criteria).
The $\chi^2_{pdf}$ cuts remove noisy, sparse light curves where the
variability is at low signal-to-noise ratio (SNR; e.g., see
Figure~\ref{fig:lceg}), rather than removing quasars that do not intrinsically
vary,  since essentially all quasars should be variable in the absence
of poor photometry or systematics \citep{but11}. 

For the $\chi^2_{pdf}$ calculation, all primary and secondary SDSS
photometric observations  are considered, along with PS1 PV2 data,  matched to within 1$''$ and without regard to  morphology or  data quality flags. The PS1 data
include observations up through December 2013, and the error inflations derived in \cite{mor15} are applied. 
Point source (PSF) magnitudes are adopted, as we are
interested in the nuclear variability  (e.g., in the case of resolved AGN).
Before the $\chi^2_{pdf}$ calculation, the SDSS magnitudes are 
transformed to the PS1 system as described in \cite{mor15}.  We also
first remove deviant points, as defined by being $>0.5$~mag from either the SDSS or PS1 running
average.  For this outlier rejection, we consider SDSS  and PS1 data separately due to
the gap in time between the two data sets. The outlier rejection is only applied to SDSS light curves with
$n\geq 10$ points (i.e., Stripe 82 data), and to PS1 data with $n\geq 5$ points.   
To compute
the running averages, we use a window of five points for SDSS data and three points
for PS1 data.  
The outlier rejection affects 5\% (10\%) of the PS1 (Stripe 82) light curves. Among these, 12.5\% (7\%) of the PS1  (Stripe 82) epochs on average
   are removed as a result.\footnote{While this criterion could lead to the rejection of
  some interesting variables,  we are mainly interested in maximizing the sample efficiency by rejecting spurious data points
  that may otherwise lead to a misleading variability measure.}

Figure~\ref{fig:lceg} shows example light curves for four sources with
S82 and PS1 photometry and with different values of $\chi^2_{pdf}$.

\subsection{Spectroscopic Data}
\label{sec:data:spec}

All samples are constructed based on known spectroscopic
classifications in SDSS. The BOSS spectrographs and their SDSS
predecessors are described in detail by \citet{smee13}. 
SDSS-III BOSS \citep{Dawson13} significantly expanded the
coverage in the SGC (approximately $-2^{\circ}<\delta<35^{\circ}$, $-30^{\circ}<\alpha<30^{\circ}$, see
Figure~\ref{fig:skycover}), and revisited the entire NGC area. 
Since the FES targets were planned before the start of SDSS-IV, 
they were restricted to SDSS-I/II/III observations. 
SDSS-IV eBOSS observations are planned to cover the entire SGC BOSS
footprint and about half the NGC, targeting mostly quasars and
galaxies \citep{blanton17}. Since the RQS targets were compiled
in summer 2016, they also draw from newly confirmed quasars targeted
as part of SDSS-IV \citep{myers15,pal16}.  
 The ELG survey \citep{raichoor17} footprint\footnote{Note: the ELG footprint has been updated
  since the version shown in Figure~1 of \cite{mor15}.}  covers most of S82, in particular the ``Thin82''
and ``Thick82'' regions outlined in Figure~\ref{fig:skycover},
totalling  620~deg$^2$ in the SGC.  
This area is also covered by eBOSS plates designed for luminous
red galaxies \citep[LRG;][]{prakash16} and quasar targets (including some that were
observed after the SDSS-IV observations shown in Figure~\ref{fig:skycover}).
The ELG footprint also includes 600~deg$^2$ in the NGC.

The wavelength coverage of the SDSS (BOSS) spectrographs is
3800--9200\AA\ (3600--10,400\AA), with a spectral resolution ranging from 1850 to
2200 (1560 to 2650). The SDSS and TDSS spectra presented in this work all have
$\lambda_{\rm eff} = 5400$\AA, i.e., the plate holes were drilled to
maximize the SNR at $\lambda_{\rm eff}$,  and the BOSS spectra 
either have $\lambda_{\rm eff} = 5400$\AA\ or $\lambda_{\rm eff} = 4000$\AA\
\citep{Dawson13}. To accurately compare spectra with differing
$\lambda_{\rm eff}$, one must correct the spectra using the
prescriptions given in \citet{margala16,guo16,harris16}.  Note that the \citet{margala16} corrections are applied in the DR14 release of spectra from the BOSS spectrographs \citep{abolfathi17}.

\subsubsection{Quasar Catalogs and Temporal Baselines}

As described by \citet{ric02}, the bulk of SDSS quasar target
candidates in SDSS-I/II were selected 
for spectroscopic observations based on their optical colors and 
magnitudes in the SDSS imaging data or their detection in the FIRST 
radio survey \citep{Becker95}. Low-redshift, $z\lesssim 3$, quasar targets
were selected based on their location in $ugri$-color space and the
quasar candidates passing the $ugri$-color selection were selected to a
flux limit of $i = 19.1$. High-redshift ($z\gtrsim 3$) objects were
selected in $griz$-color space and are targeted to $i = 20.2$.
Furthermore, if an unresolved, $i \leq 19.1$ SDSS object was matched to
within 2$''$ of a source in the FIRST catalog, it was included in the
quasar selection. Additional quasars were also (inhomogeneously) discovered and cataloged 
in SDSS-I/II using X-ray, radio, and/or alternate odd-color information, 
and extending to fiber-magnitudes of about $m<20.5$ \citep[e.g., see][]{anderson03}.

\begin{figure}
\epsscale{.7}
\plotone{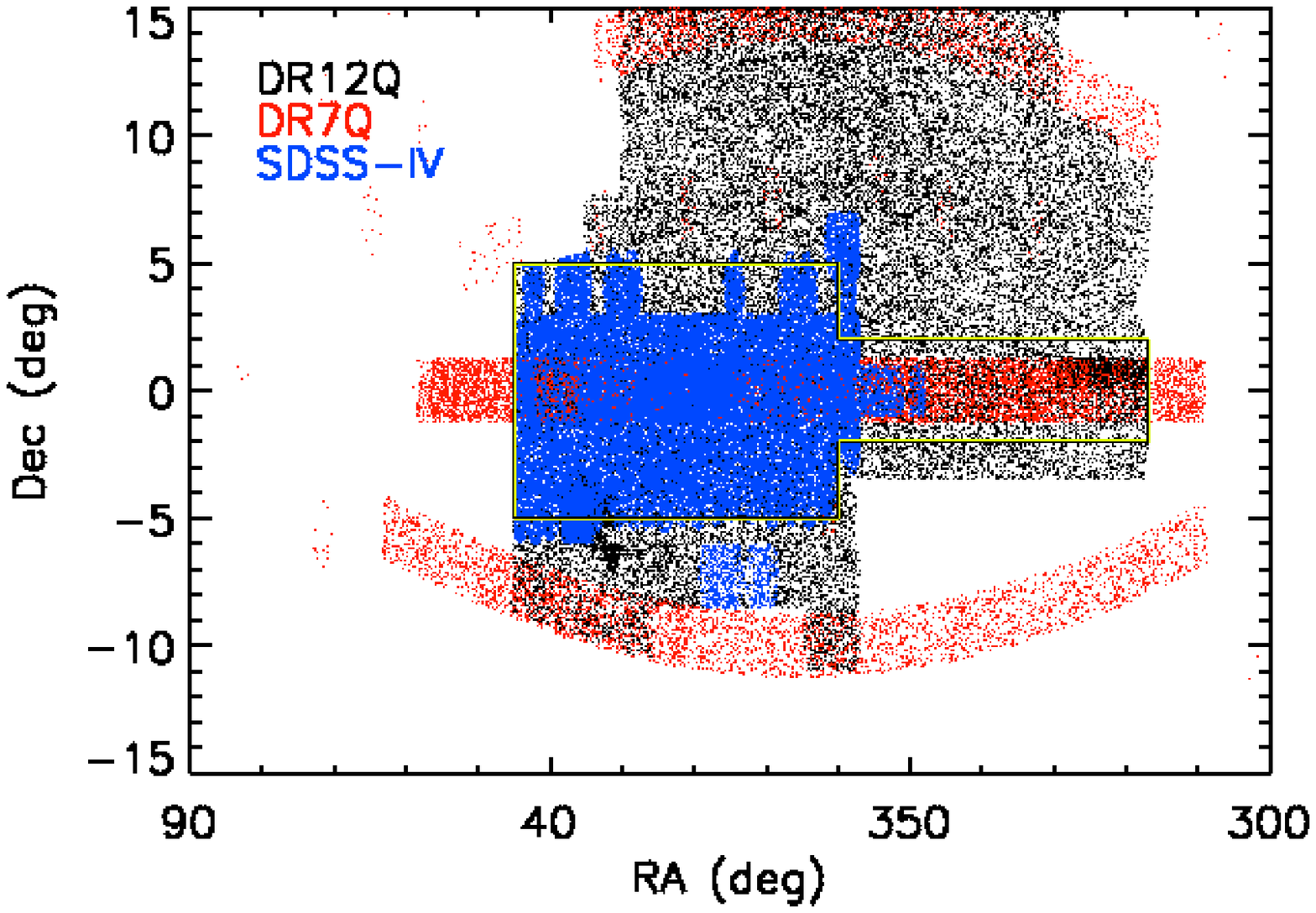}
\caption{Sky coverage in the SGC (J2000 coordinates) as of early 2016 (SpAll version 
  v5\_9\_1). The ``Thin82'' and ``Thick82'' chunks are outlined, with
  Thin82 covering a subset of $315^{\circ}<\alpha<360^{\circ}$, $-2.0^{\circ} < \delta <
  2.75^{\circ}$, and Thick82 covering $0^{\circ}<\alpha<45^{\circ}$, $-5^{\circ} < \delta <
  5^{\circ}$.
 \label{fig:skycover}}
\end{figure}

Unless otherwise stated, we select quasars for repeat spectroscopy
from one of the visually vetted quasar catalogs: 
the SDSS-I/II DR5/7 quasar catalogs \citep[DR5Q,
DR7Q;][]{schn07,sch10,shen11} or the DR12 quasar catalog \citep[DR12Q,
final quasar catalog of SDSS-III;][]{par16}. 
The RQS target selection considers confirmed SDSS-IV
quasars from post-DR13 data \citep{myers15,pal16}, specifically the
{\tt SpAll} database version {\tt v5\_9\_1}, which covers a region in the SGC (see 
Figure~\ref{fig:skycover} and Table~\ref{tab:sdss4} for the SDSS-IV
coverage at the time of target selection).
 The following SDSS-IV spectra are excluded from consideration:
\begin{enumerate}
\item  Those objects lacking primary ({\tt mode}$=1$)
  magnitudes in DR10 (these objects are faint and few in number);
\item  Those with {\tt OBJTYPE} $=$ {\tt SKY};
\item  Those objects with morphological {\tt TYPE} $= 0$, according to DR10 photometry; and
\item  Those at redshift $z\geq 0.8$  with morphological {\tt TYPE} $=
  3$, since a resolved quasar should be at a lower redshift.
\end{enumerate}
\noindent In order to determine the number of existing spectra in
\S\ref{sec:rqs}, we extract all spectroscopy within 2$''$  from the 
SDSS DR12 {\tt SpecObjAll} database. For the new 
SDSS-IV objects, we use the {\tt NSPEC} field from {\tt SpAll-v5\_9\_1}.

\begin{figure}
\plotone{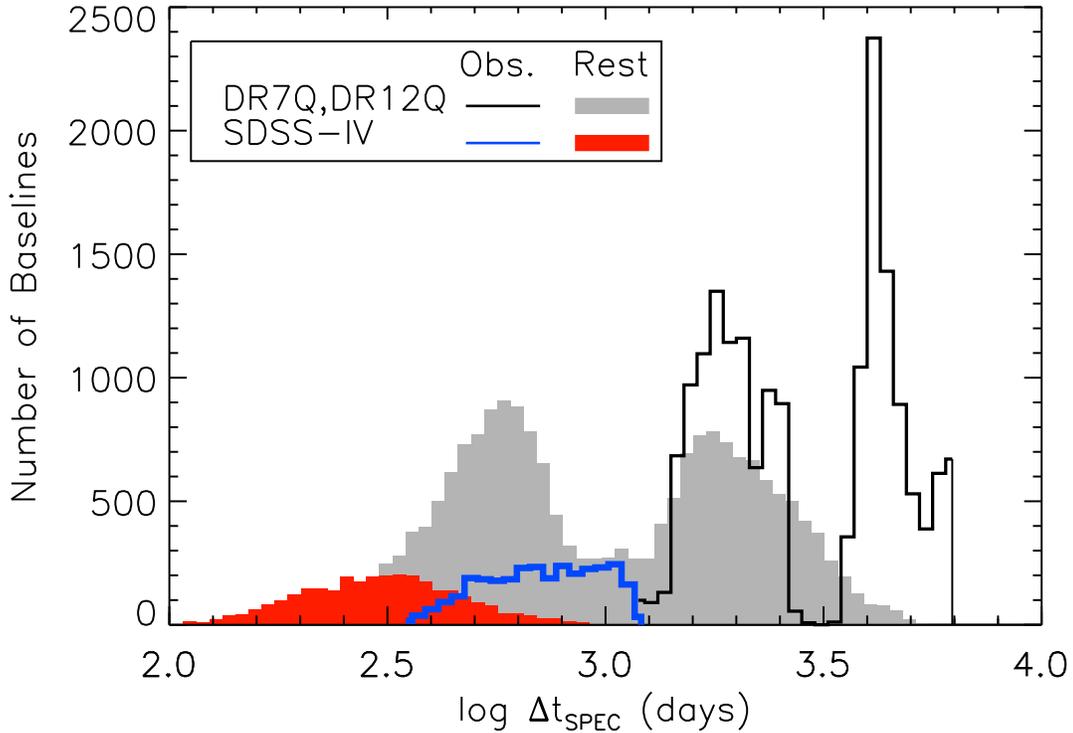}
\caption{Distribution of anticipated time baselines in the observed 
  and rest frames (open and filled histograms, respectively) for quasars
  considered for repeat spectroscopy in the SDSS-IV RQS program (\S\ref{sec:rqs}).
  The RQS epochs are
  artificially set to a uniform distribution over the year 2017.  
  The total distribution of $\sim$20,000 quasar baselines in the
  observed (rest) frame is shown as the open black
  (filled gray) histogram, where a single existing spectroscopic epoch is adopted.  
  The baselines in the observed (rest) frame for objects with existing SDSS-IV spectra ($\sim$3,000
  total) are shown as the open blue (filled red) histogram. 
\label{fig:timescales}}
\end{figure}

\begin{figure}
\plotone{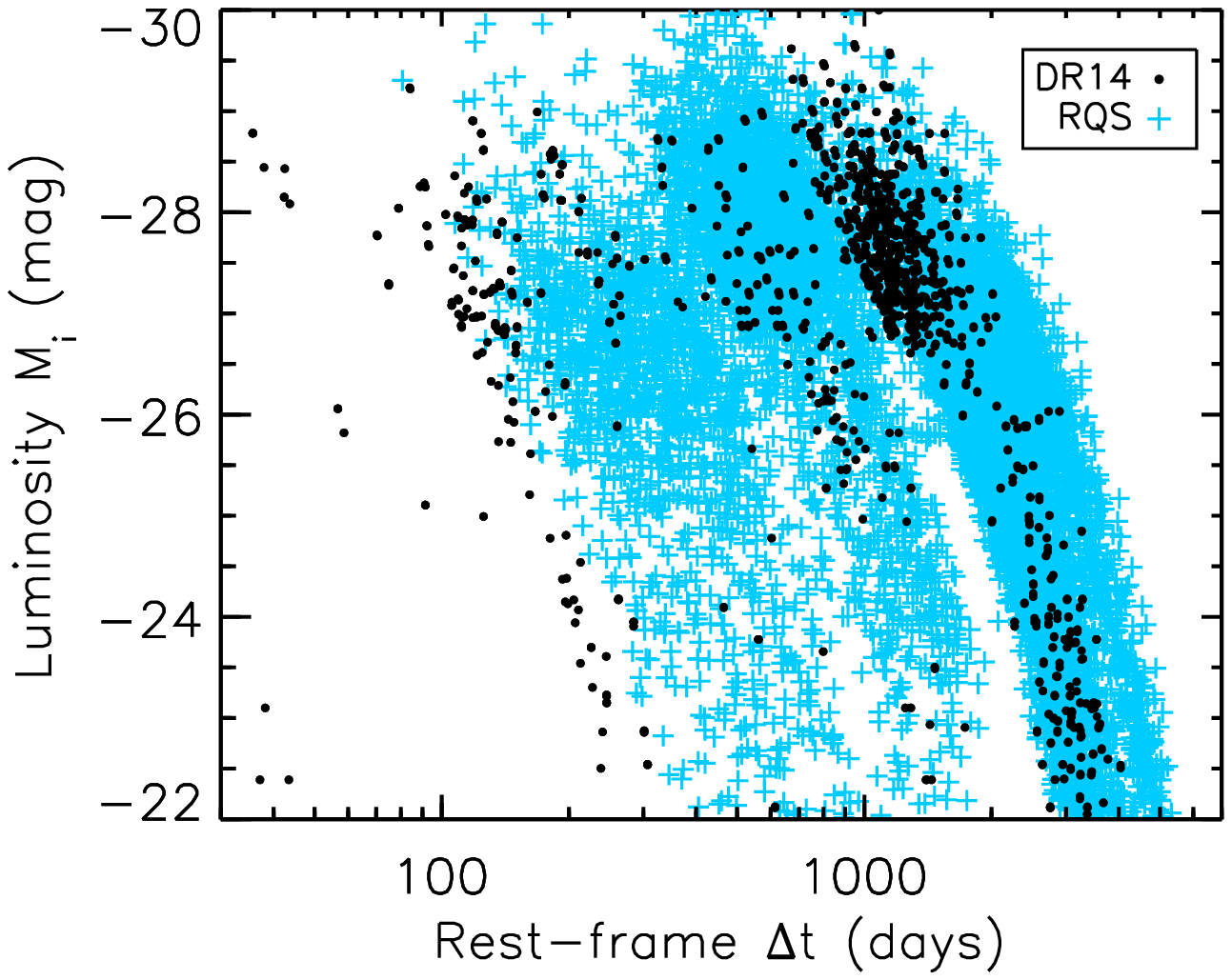}
\caption{The distribution of rest-frame time intervals between spectra
  as a function of luminosity for $i<19$ quasars in DR14  
  is shown as black dots. The points are restricted to quasars
  within a representative area of 233~deg$^2$
  that have more than one existing spectroscopic epoch, and include
  all existing pairs of epochs  in DR14. 
  The anticipated distribution for $i<19$ quasars targeted by the
  RQS program (\S\ref{sec:rqs}) is shown in cyan (adopting a single
  existing epoch for each quasar).  The RQS epochs are
  artificially set to a uniform distribution over the year 2017.  
\label{fig:dtmi}} 
\end{figure}

Figures \ref{fig:timescales} and \ref{fig:dtmi} show the anticipated 
distribution of time lags between spectra for SDSS quasars.  
Figure~\ref{fig:dtmi} also shows the existing distribution of time
lags and displays these distributions as a function of  absolute
magnitude $M_i$, where the $M_i$ values are estimated from the apparent
$i$ magnitudes and distance modulus (no K-correction is applied).
Note that these Figures do not include the well-sampled quasar cadence from the
SDSS Reverberation Mapping Program \citep{shen15}.

\section{Few-Epoch Spectroscopy (FES)}
\label{sec:fes}
In addition to its main program to obtain initial characterization spectra of
$>10^5$ optical variables selected from PS1, the TDSS
includes nine separate smaller, FES programs
to study spectroscopic variability. The FES programs target objects
with existing SDSS spectroscopy amongst classes of quasars and stars
of particular astrophysical interest to build statistical samples for
follow-up study.\footnote{Note that the FES program approach is 
conceptually somewhat different than, and complementary to, the RQS 
approach. The RQS intentionally -- and with fewer a priori biases -- 
samples spectral variability across a much broader range of quasar 
subclasses, whereas the FES subclasses are more-specific-science focused 
(e.g., BALQSOs), and therefore efficiently address some more 
restricted questions. The FES subclasses are custom-tuned and relatively 
smaller than the RQS, but still large in sample-size  in an 
absolute sense with hundreds to thousands each, providing 
excellent statistics albeit attuned to more highly- and 
specifically-selected subsamples.} These include, in approximate
order of decreasing sample size: BAL quasars, the most photometrically variable (``hypervariable'')
quasars, high SNR normal broad-line quasars, quasars
with double-peaked or very asymmetric broad emission line profiles,
hypervariable stars (including the most highly variable classical pulsators), active ultracool (late-M and early-L) dwarf stars
with H$\alpha$ emission, dwarf carbon stars, white dwarf/M dwarf
spectroscopic binaries with H$\alpha$ emission, and binary
supermassive black hole candidates from \ion{Mg}{2} broad line velocity shift
analysis.  

The FES programs and respective scientific goals and target selections
are described  in the following sections, starting with stars
(\S\ref{sec:fes:ACTSTAR}-\ref{sec:fes:HYP}) and ending with quasars
(\S\ref{sec:fes:HYP}-\ref{sec:fes:NQHISN}). The target flags used are listed
in each subsection heading.

\subsection{Magnetic activity on late-M and early-L dwarfs {\tt (TDSS\_FES\_ACTSTAR)}} 
\label{sec:fes:ACTSTAR}
Magnetic activity is ubiquitous in stars at the transition between the
M and L spectral types (ML dwarfs). In optical spectra, this activity
is best identified with the H$\alpha$ emission line, which traces
chromospheric heating on these low mass objects
\citep[e.g.,][]{Gizis2002a,west11}. H$\alpha$ is the optimal
diagnostic in part because it is found in a relatively red portion of
the spectrum, so is easier to observe for these very cool, red
objects. Serendipitous and dedicated observations of H$\alpha$ emission on multiple timescales have indicated H$\alpha$ emission varies \citep[sometimes dramatically;][]{Hall2002} on multiple timescales \citep[e.g.,][]{Berger2009,sch15}. Chromospheric heating covers only a small portion of the surface \citep[$<$1\%][]{sch15}, and those regions rotate in and out of view on timescales of hours to days \citep[due to relatively rapid rotation;][]{Reiners2008} leading to H$\alpha$ variability. On timescales of weeks to months, we expect the chromospheric emission regions to change in size due to underlying shifts in the magnetic field (similar to shifts in sunspots).  H$\alpha$ may also show variability over year to decade-long timescales based on long-timescale magnetic field changes that are similar to the 11-year solar cycle.  Analyses of H$\alpha$ variability on ML dwarfs have so far been limited to $\sim$20 objects serendipitously observed by multiple groups, but the data indicate that 30–-50\% of ML dwarfs exhibit significant variability over timescales that span months to years \citep{sch15}. 

By comparing original spectra of ML dwarfs from the SDSS legacy survey
with an additional spectrum from TDSS, we have a unique opportunity to
monitor changes in H$\alpha$ emission lines over timescales of 6--14
years. These observations can either be taken as indicators of the
level of overall variability, or could be combined with data over
shorter timescales to detect decadal magnetic cycles. The SDSS data for ML dwarfs also allows three-dimensional Galactic kinematics that can be leveraged to examine age and activity correlations among the multi-epoch observations. 

To select the FES sample of ML dwarfs, we combined the \citet{west11}
M dwarf and \citet{Schm10} L dwarf catalogs. We selected a subset of dwarfs from those catalogs with magnitudes between $17 < i < 21$ and spectral types from M7 to L3. We also
required an average SNR $> 3$ per 1.5\AA\ pixel in the continuum surrounding H$\alpha$
(6530--6555\AA\ and 6575--6600\AA) so that  the presence and strength of H$\alpha$ emission can be reliably measured \citep[e.g.,][]{west08}. The dwarfs in our initial sample are contained in the BOSS Ultracool Dwarfs catalog (Schmidt et al. 2017, in prep.), and we required that each of the dwarfs have a photometric distance (based on SDSS photometry), proper motion (from SDSS-2MASS-WISE positions) and radial velocity (based on SDSS spectroscopy) from that catalog. We also restricted the sample to dwarfs within 300~pc of the Galactic plane. 

\begin{figure}
\epsscale{.7}
\plotone{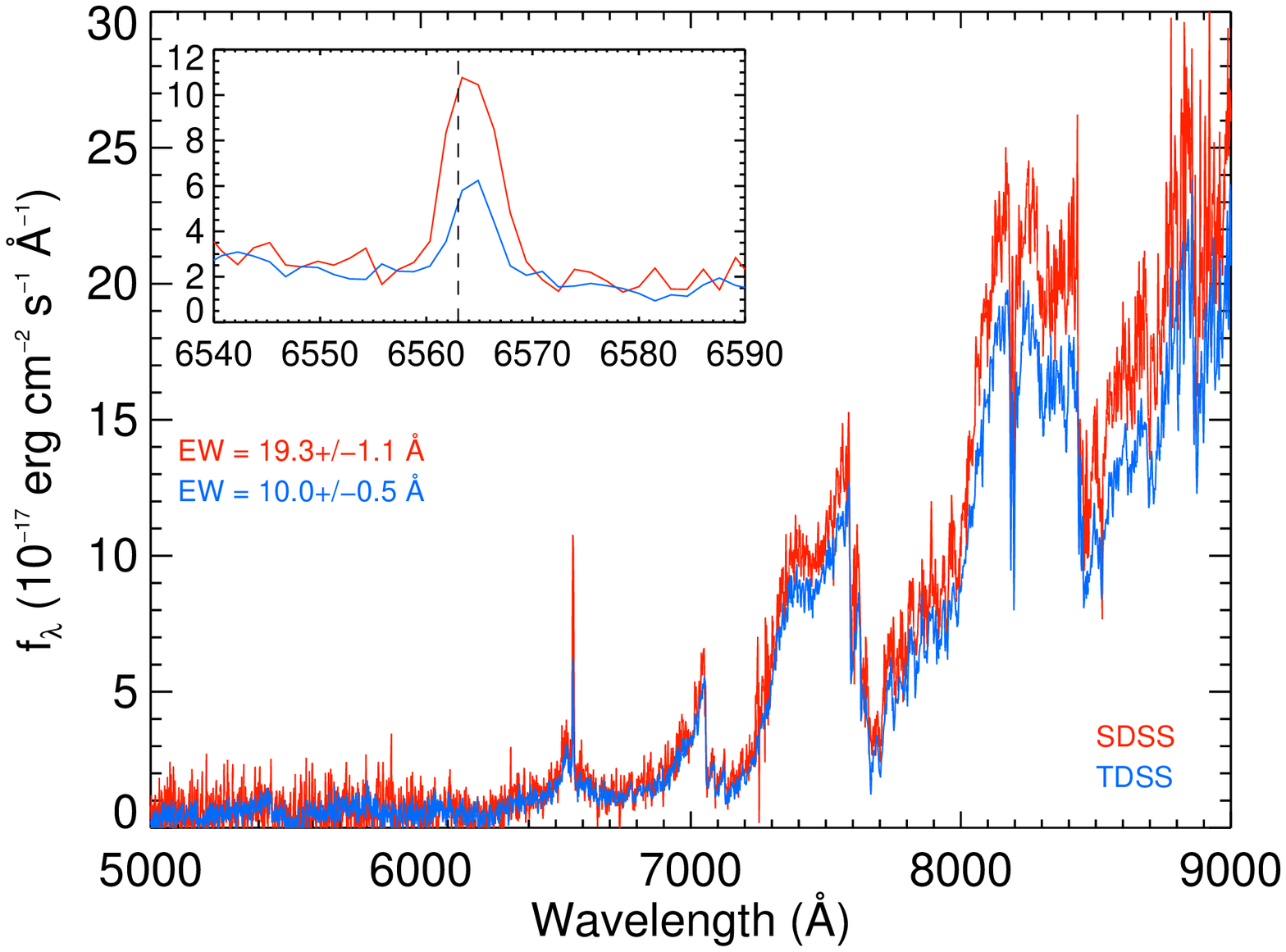}
\plotone{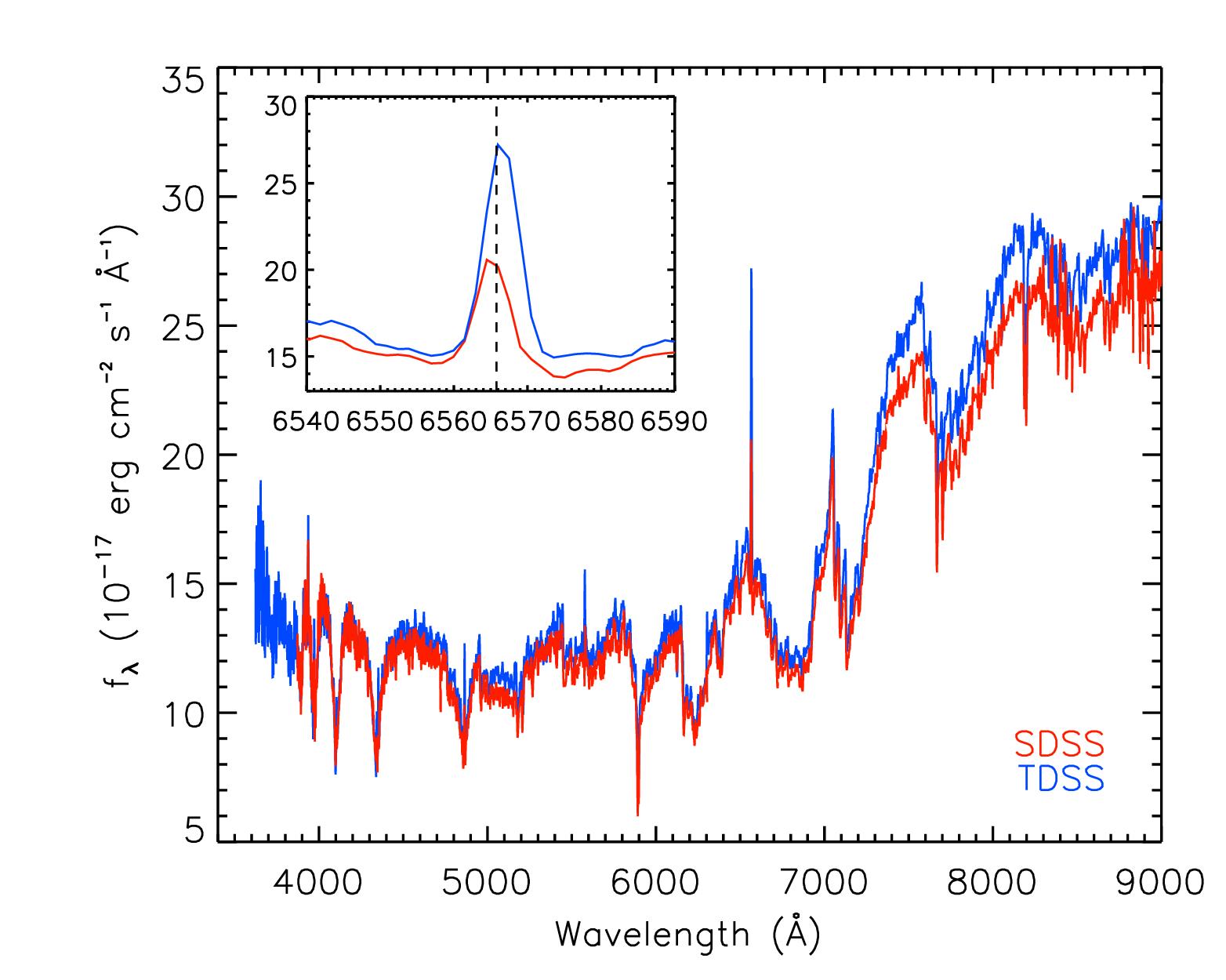}
\caption{{\it Top:} SDSSJ~014724.37+005751.4, an M9 dwarf (target class {\tt
    TDSS\_FES\_ACTSTAR}, see
  \S\ref{sec:fes:ACTSTAR}) first
  observed with the original SDSS spectrograph (black) with an
  additional epoch of spectroscopy from TDSS (red). The inset displays
  the area surrounding the H$\alpha$ emission line. This object has
  strong variable H$\alpha$ emission; between the
  original epoch (MJD 51793) and the more recent epoch (MJD 56902) the
  emission line weakened significantly. It is not yet known whether
  this behavior reflects variations on a short timescale or longer
  timescale evolution of the magnetic field. 
  {\it Bottom:} SDSSJ~231105.67+220208.7, a WD-dM binary  (target class {\tt
    TDSS\_FES\_WDDM}, see \S\ref{sec:fes:WDDM}) that shows a
  significant brightening of H$\alpha$ in the TDSS spectrum over 7.9~yr (see
  inset). This object has a binary period of 13.9 hours
  \citep{nebot11}. The same level of smoothing has been applied to both
  SDSS and TDSS spectra, and the observed frame is shown.
 \label{fig:MLdwarf}}
\end{figure}

The selection criteria resulted in a total of 3739 M7, 534 M8, 153 M9, and 23 L
dwarfs. To reduce the sample to $\sim$1000 ML dwarfs, we binned the
data by height above the Galactic plane and restricted each 25~pc-wide bin to 60
dwarfs randomly drawn from each spectral type. The final target list
includes 1036 stars (583 M7, 283 M8, 147 M9, and 23 L dwarfs). Initial data
from the ML dwarf FES sample includes dwarfs that have no change in
their activity level as well as those that have strong
variability. The spectrum of a strongly active and variable M dwarf is shown in Figure~\ref{fig:MLdwarf}. 

\subsection{ White Dwarf-M Dwarf binaries {\tt (TDSS\_FES\_WDDM)}} 
\label{sec:fes:WDDM}
Recent studies have demonstrated that close binaries consisting of
white dwarf-M dwarf pairs (WD-dM binaries) have a
significant effect on the magnetic activity of their main sequence components
\citep[e.g.,][]{mor12}. 
The binary separations where increased magnetic
activity is observed extend to $\sim$100~AU. While many of the WD+dM
are unresolved photometrically, 
the two components can be separated in low-resolution spectroscopy due
to their vastly different spectral energy distributions.
While there is evidence of increased magnetic activity in close pairs,
there has been limited analysis of the variability of magnetic
activity in binary systems. Variability studies can distinguish among
possible causes of activity (e.g., irradiation, accretion, disk
disruption, and spin-up).
This program aims to re-observe $\sim$400 WD-dM binaries identified via their
spectral energy distributions. By measuring the magnetic activity of
the M dwarf via the H$\alpha$ equivalent width (EW), the goals are to determine
\emph{(i)} the effect of binary separation on the variability of
  magnetic activity,  
\emph{(ii)} the effect of rotation on stellar
  activity in close binaries, and 
\emph{(iii)} the WD cooling age, spectral type, orbital parameters,  
  metallicity, and Galactic height, and the corresponding effects on magnetic activity.

Starting with the WD-dM binary sample of \citet{mor12}, targets that
met the following criteria were selected:
\begin{enumerate}
\item Within the magnitude range $17 < i < 21$, 
\item Clear H$\alpha$ emission from visual inspection,\footnote{
We found that simple EW and SNR criteria with fixed
wavelength intervals led to unreliable results if there were significant
shifts in line location due to orbital motion of the close binaries.
Preliminary EW and SNR values were therefore checked visually.} and
\item Accurate proper motion measurements from the SDSS-USNOB {\tt proper motions}
  table ({\tt MATCH}$ =1$; {\tt PMRA} and {\tt
    PMDEC}$\neq 0$; and {\tt DIST22}$ > 7''$, where {\tt DIST22} is
  the distance to the nearest neighbor with $g < 22$). This criterion
  is necessary in order for the fiber holes to be drilled in the
  correct locations.

\end{enumerate}
The resulting sample contains 402 active, WD-dM pairs that span several M dwarf spectral types. 
 An example
 of a binary with variable H$\alpha$ is shown in the bottom panel of Figure~\ref{fig:MLdwarf}.

\subsection{ Variability in Dwarf Carbon Stars {\tt (TDSS\_FES\_DWARFC)}} 
\label{sec:fes:DWARFC}
Carbon in stellar atmospheres - indeed, most of the carbon in the
universe - is produced by the triple-$\alpha$ process of helium fusion
(3\,$^4$He\,$\rightarrow\,^{12}$C) in the interiors of red giant
stars. Strong carbon molecular bands are historically expected to be
seen only in  asymptotic giant branch (AGB) stars that have
experienced a third ``dredge-up'' \citep{iben83}.  However, among stars showing such C$_2$ and CN molecular bands (C stars), the main sequence carbon dwarfs (dCs) are numerically
dominant in the Galaxy \citep{gre92}.  The accepted explanation for dCs is that they must all be in
post-mass transfer binaries, where the former AGB star has since
become a white dwarf, leaving a carbon-enhanced dC primary. Indeed,
a handful of ``smoking gun'' systems reveal evidence for this evolutionary scenario, having
composite spectra with a hot DA white dwarf component \citep{heb93,lie94,gre13,si14}.
While the connection has rarely been made in the literature, dC stars, having been rejuvenated by
mass accretion, would likely be seen as blue stragglers if they were
within a co-eval stellar cluster.  They are also probably the dwarf progenitors of the typically more luminous carbon-enhanced metal-poor (CEMP), sgCH, CH and perhaps barium (Ba II) stars, which all show carbon and $s$-process enhancements (see discussion and references in \citealt{demarco17}).  The detection of white dwarf companions, and the characterization of the orbital properties of dC stars, is therefore important for understanding the mass transfer processes that give rise to this fascinating family of stars.

However, to date, the only dC star with a measured binary orbit is the
prototype dC G77-61 \citep{dah77,dea86}, a single-line
spectroscopic binary (245 day period and semi-amplitude 20\,km~s$^{-1}$), where the WD
has cooled to $T_{\rm eff}<$ 6000\,K.  Since G77-61 represents the
only known dC with a proven radial velocity (RV) orbit, the mass-transfer hypothesis for dCs remains to be confirmed, 
and can be investigated only through the properties of dCs to detect and characterize  host binary
systems.  Models for dC formation in both the disk and halo \citep{dek95} predict a bimodal orbital period distribution, with a large peak at a $\sim$decade (for accretion of the AGB wind at a binary
separation $\sim$10\,AU), and a smaller peak at $\sim$year (for
separations $\lesssim 1$\,AU) corresponding to systems that underwent a
common envelope (CE) phase, where the companion was subsumed in the
expanding atmosphere of the AGB star when it fills its Roche lobe.
These models reproduce the better-studied distributions of CH and
Ba\,II giants, whose progenitors are almost certainly the dCs.  The relic distribution of dC binary orbits 
should reveal the relative importance and efficiency of these types of accretion, which can substantially modify the dC, leaving it hotter and bluer (and perhaps more rapidly rotating) than expected for its age.

\citet{gre13} identified 1220 faint ($r\gtrsim 17-21$) C stars from
SDSS spectra, $\sim$5 times more than previously known, but also including a
wider variety of dC properties than past techniques such as color or grism selection
have netted.  From those with significant proper motion measurements, they
identified 730 definite dwarfs, including eight systems with clear
DA white dwarf companions. This dataset represents the first significant sample of
{\em bona fide} dCs appropriate for a population study.

The statistical analysis of large samples of sparsely sampled RV
curves can be used to constrain the underlying properties (binary
fraction and separation distribution) of the corresponding binary
population (e.g., \citealt{mao12}).   
The TDSS dwarf carbon star FES program will provide a second epoch of SDSS
spectroscopy to measure RV variability for a large sample
of dC stars, to produce first constraints on their binarity
and the distribution of their orbital properties. The main aims of
this program are to 
\emph{(i)} test the binary evolution hypothesis for dwarf Carbon (dC) stars,
\emph{(ii)} constrain the distribution of orbital separations, and 
\emph{(iii)} trace the chemistry and evolution of the oldest asymptotic-giant-branch (AGB) stars.
The strategy used in the program will: \emph{(i)} measure the RV shift
$\Delta RV$ for dC stars between SDSS and TDSS (5--18
years), and  \emph{(ii)} constrain the \emph{separation} distributions and hence the mass transfer mode.

\begin{figure}
\plotone{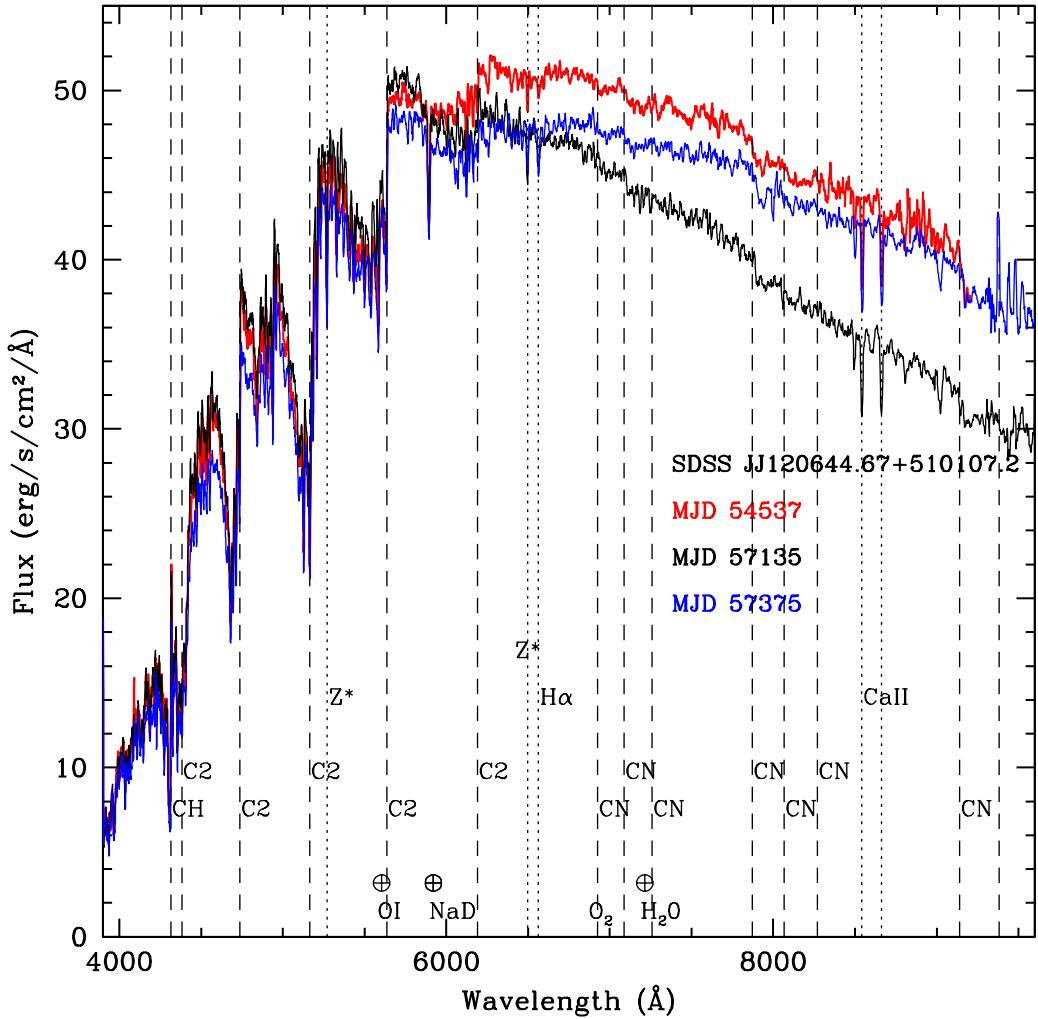}
\caption{Example dwarf carbon star from TDSS (target class {\tt TDSS\_FES\_DWARFC}, \S\ref{sec:fes:DWARFC}). 
The spectroscopic MJDS are 54537, 57135, and 57375.  Prominent molecular bandheads of CH, C$_2$, and CN are labeled and marked by vertical dashed lines.  Several other atomic features are marked with vertical dotted lines, including strong atomic metal line blends (Z$^*$), H$\alpha$, and Ca\,II.   The location of telluric features are marked across the bottom.  TDSS allows study of RV variability, changes in color and brightness, and line strengths to illuminate the physics of these unique post mass transfer binaries. 
\label{fig:dCspec}}
\end{figure}

For the dC FES program, we selected all 730 SDSS C stars from
\citet{gre13} that were listed as dwarfs with high probability
based on either their measured proper motions, or because they were 
identified from their SDSS spectra as composite DA/dC spectroscopic
binaries.  We added another 99 dC stars found by \citet{si14}, totalling 829 unique dC stars for repeat spectroscopy within TDSS.
An example of SDSS archival and
TDSS spectra of a dC in our program is shown in Figure~\ref{fig:dCspec}.

\subsection{ ``Hypervariable'' stars and quasars {\tt
   (TDSS\_FES\_HYPSTAR}, {\tt TDSS\_FES\_HYPQSO)}} 
\label{sec:fes:HYP}
This program targets the most highly photometrically variable stars/classical pulsators (defined as hypervariable stars), as well as hypervariable 
quasars in the TDSS.  The
spectroscopic variability for these objects can potentially reveal large structural changes in
astrophysical sources, and is useful for finding rare,
transient phenomena such as ``changing-look quasars''
\citep[e.g.,][]{lam15,run16,rua16a,mac16}.  More importantly, this program is
exploring unknown territory and therefore the scientific returns 
could be quite substantial.
  
  During the variable target selection in the main TDSS SES program \citep{mor15}, 
hypervariables are identified using a modified variability characterization 
that is designed to work in the extreme regions of variability space
(see \S\ref{sec:data:var}).  
The hypervariable targets for these FES programs all have previous
spectra in the SDSS DR11 {\tt SpecObjAll} table.  Since the pipeline
classifications were adopted here without further verification, a few targeted stars may actually be
quasars, and vice versa.

\begin{figure}
\plotone{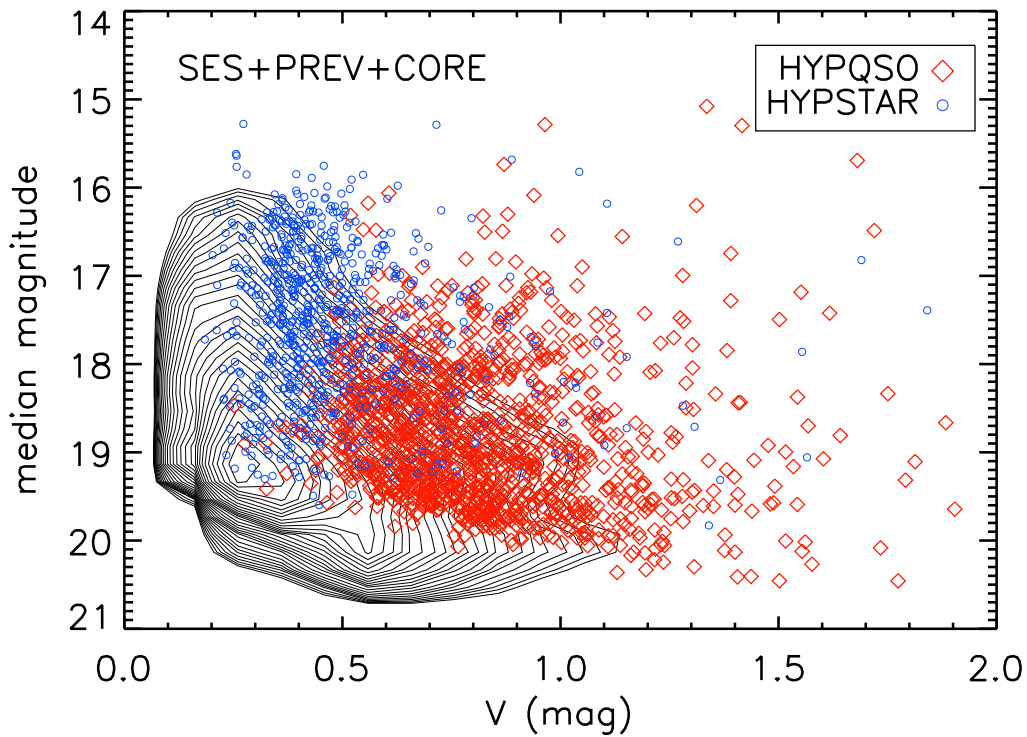}
\caption{ Distribution of the variability metric, $V$ (Eqn.~\ref{eq:V}), as a function of
  median magnitude among PS1 $griz$ filters for all
  variability-selected quasars and stars described in
  \citet{mor15}.  The logarithmic contours show the overall
  distribution for TDSS variables that either: a) are targeted for single epoch spectroscopy (SES) in TDSS, 
  b) already have pre-existing spectra in SDSS (PREV), or c) are also
  targeted as part of the eBOSS CORE quasar program.  The red diamonds
  (blue circles)
  show the distribution for hypervariable quasars (stars) targeted by
  the FES programs described in \S\ref{sec:fes:HYP}.
 \label{fig:vdist}}
\end{figure}

\begin{figure}
\epsscale{.6}
\plotone{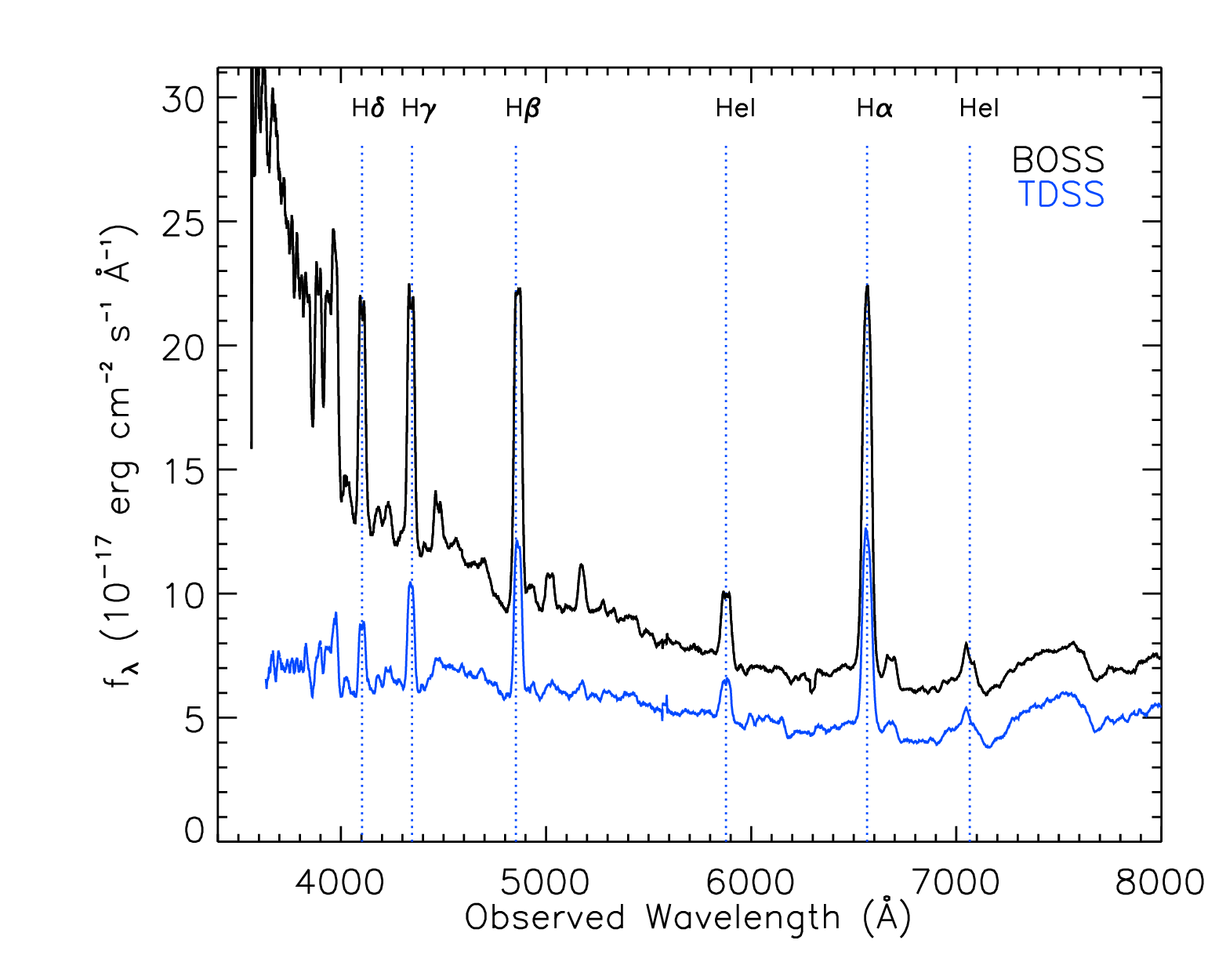}
\plotone{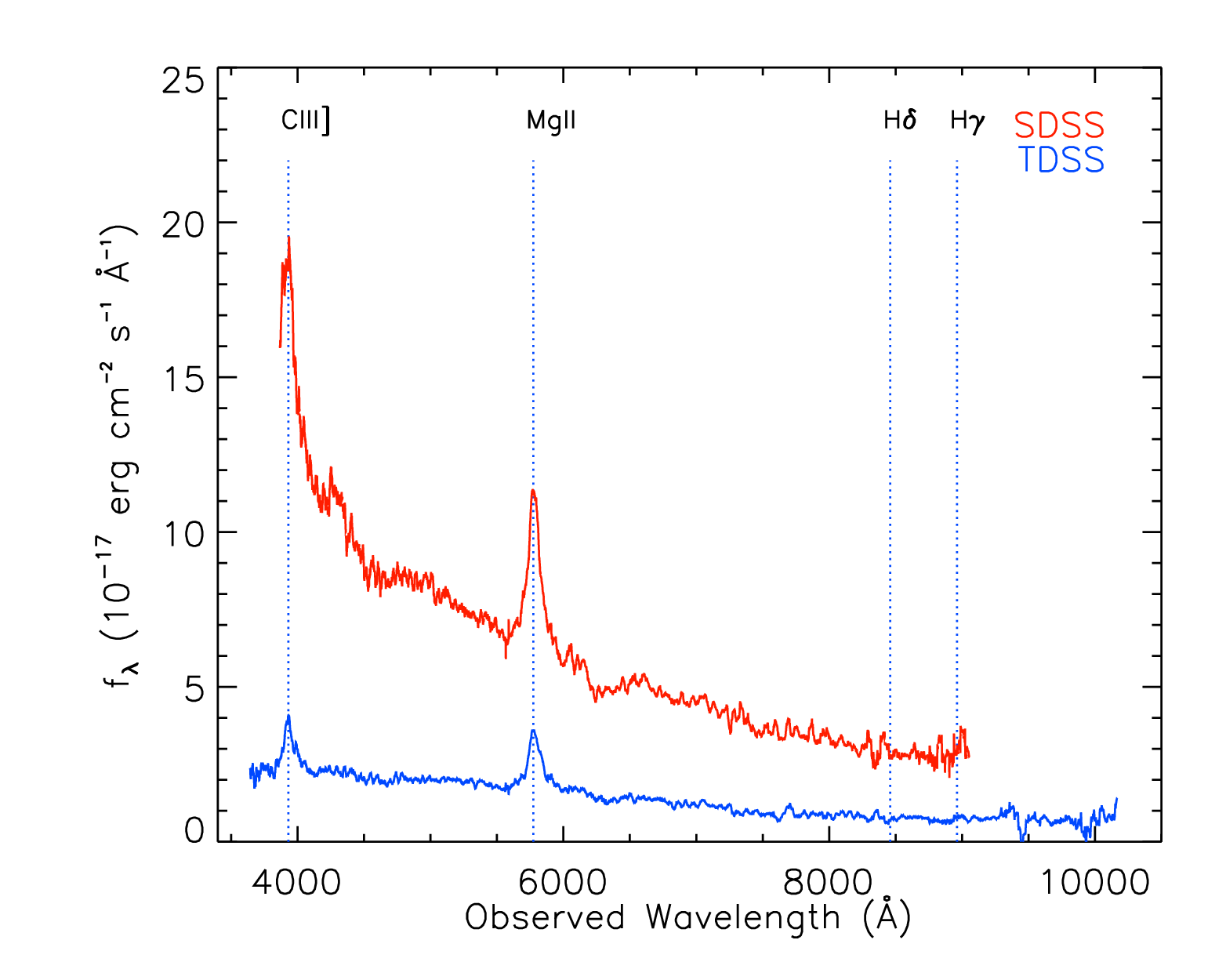}
\plotone{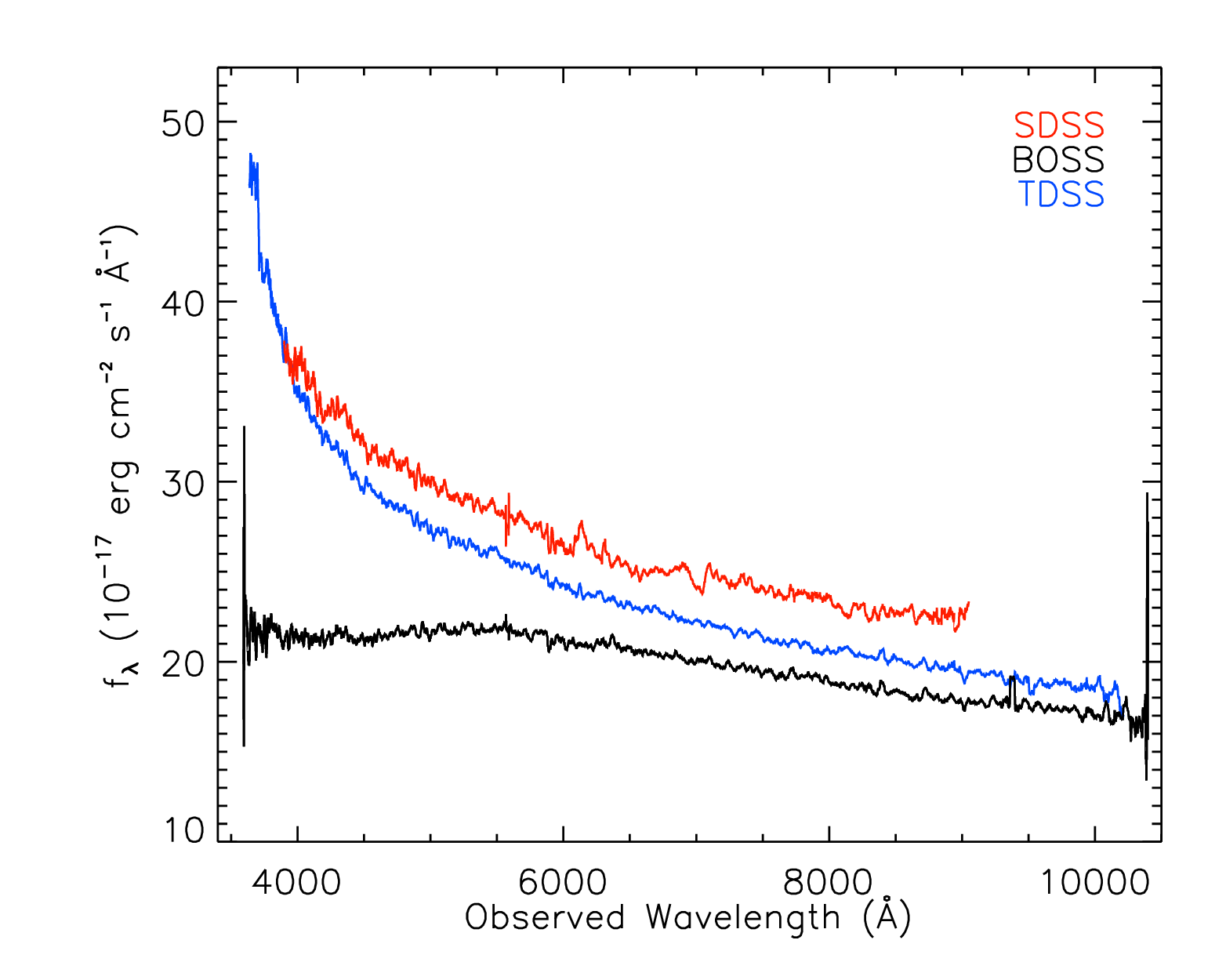}
\caption{Examples of hypervariables (\S\ref{sec:fes:HYP}) targeted by the {\tt
    TDSS\_FES\_HYPSTAR} (top panel), and {\tt TDSS\_FES\_HYPQSO}
  (bottom two panels) programs. The object in the top panel is 
  cataclysmic variable SDSS~J003827.04+250925.0; the time between
  spectra is 2.9~yr.  The middle panel shows hypervariable quasar
  SDSS~J235040.09+002558.8 at redshift $z=1.062$, exhibiting a large change over 14.2~yr
  (observed frame). Shown in the bottom panel is blazar
  SDSS~J081815.99+422245.4 also found in \citet{massaro14}, with spectroscopic
   MJDs 52205, 55505, and 57361 for SDSS, BOSS, and TDSS, respectively. 
 \label{fig:hypeg}}
\end{figure}

For stars, defined as $i< 20$ point sources (uncorrected for Galactic extinction) with 
{\tt CLASS} $=$ {\tt STAR}, the top 0.5\%
most significantly variable objects were selected, corresponding
approximately to $V>0.3$~mag (see Figure~\ref{fig:vdist}). 
These sources lie outside an approximately elliptical contour with SDSS-PS1
difference of 0.2~mag, a PS1-only variability of 0.15~mag, 
or some intermediate combination of the two \citep[c.f.\ Figure~5,][]{mor15}.  The SDSS images of these
sources are visually examined to remove objects with close
neighbors, nearby diffraction spikes or other imaging issues that
could significantly affect photometry.
The above criteria select 1150 stars  ($\sim$0.05~deg$^{-2}$), which
have the target flag {\tt TDSS\_FES\_HYPSTAR}.  Inspection of these targets'
initial SDSS spectra suggest that this sample is rich in RR Lyrae
variables and
also includes M dwarfs, carbon stars and stars that are difficult to
classify. For an example target, see Figure~\ref{fig:hypeg}.

For quasars, defined as $i< 20$ point sources  (uncorrected for Galactic extinction) with 
{\tt CLASS} $=$ {\tt QSO}, the top 2\% most
significantly variable objects were selected, corresponding to approximately $V>0.5$~mag (see Figure~\ref{fig:vdist}). These
sources lie outside an approximately elliptical contour with SDSS-PS1
difference of 0.7~mag, a PS1-only variability of 0.25~mag,
or some intermediate combination of the two. The SDSS images of these
sources are also visually examined to remove objects with close neighbors,
nearby diffraction spikes or other imaging issues that could
significantly affect photometry. 
The above criteria select 1555 quasars ($\sim$0.05~deg$^{-2}$), which have 
the target flag {\tt TDSS\_FES\_HYPQSO}. Inspection of these targets' initial SDSS
spectra suggest that this sample is rich in BAL 
quasars and blazars, but otherwise contains a wide range of quasar
types (we leave a detailed census to a later publication).
For example spectra, see Figure~\ref{fig:hypeg}.

\subsection{ Broad absorption line variations in quasars {\tt (TDSS\_FES\_VARBAL)}} 
\label{sec:fes:VARBAL}

This FES program will build upon recent systematic, sample-based
studies of BAL variability \citep[e.g.,][and
references therein]{barlow93,lundgren07,ak12,ak13,ak14,viv14} by re-observing $\sim$3000 BAL quasars
from SDSS and BOSS.   About $2/3$ of the sample was selected from
\citet{gib09} and has already been mostly observed as part of a BOSS
ancillary proposal \citep[see][]{ak13} and probes rest-frame timescales
of $\approx$4--7~yr. TDSS is obtaining a third 
spectroscopic epoch for this subsample, typically spanning an additional
1--3~yr in the rest frame beyond the most recent BOSS observations.  
The TDSS data  yield improved measurements of the dependence of BAL 
EW variability upon rest-frame timescale, enabling a test of the extent to which long-term variability
trends found in the SDSS-I/II vs.\ BOSS data persist.
A third epoch also allows for the possibility of detecting BAL acceleration 
or re-emergence/disappearance.   The long timescales sampled by this
project are highly beneficial since velocity shifts associated with 
BAL acceleration/deceleration accumulate over time; the first results
on this project's BAL acceleration are presented in \citet{gri16}.
Constraints upon BAL disappearance and emergence provide key
insights into the lifetime of BALs. Furthermore,  BAL re-emergence events at the same velocity argue
strongly against models where the variability is due to gas motions, 
instead favoring models where ionization changes play a key role.
The first results on BAL re-emergence/disappearance
are presented in \citet{mcgraw17}. 
Finally, these observations further characterize the coordinated EW variations of BAL 
quasars with multiple troughs; these coordinated variations constrain 
models for BAL variability \citep[e.g.,][]{ak12,ak13}.

\begin{figure}
\epsscale{.6}
\plotone{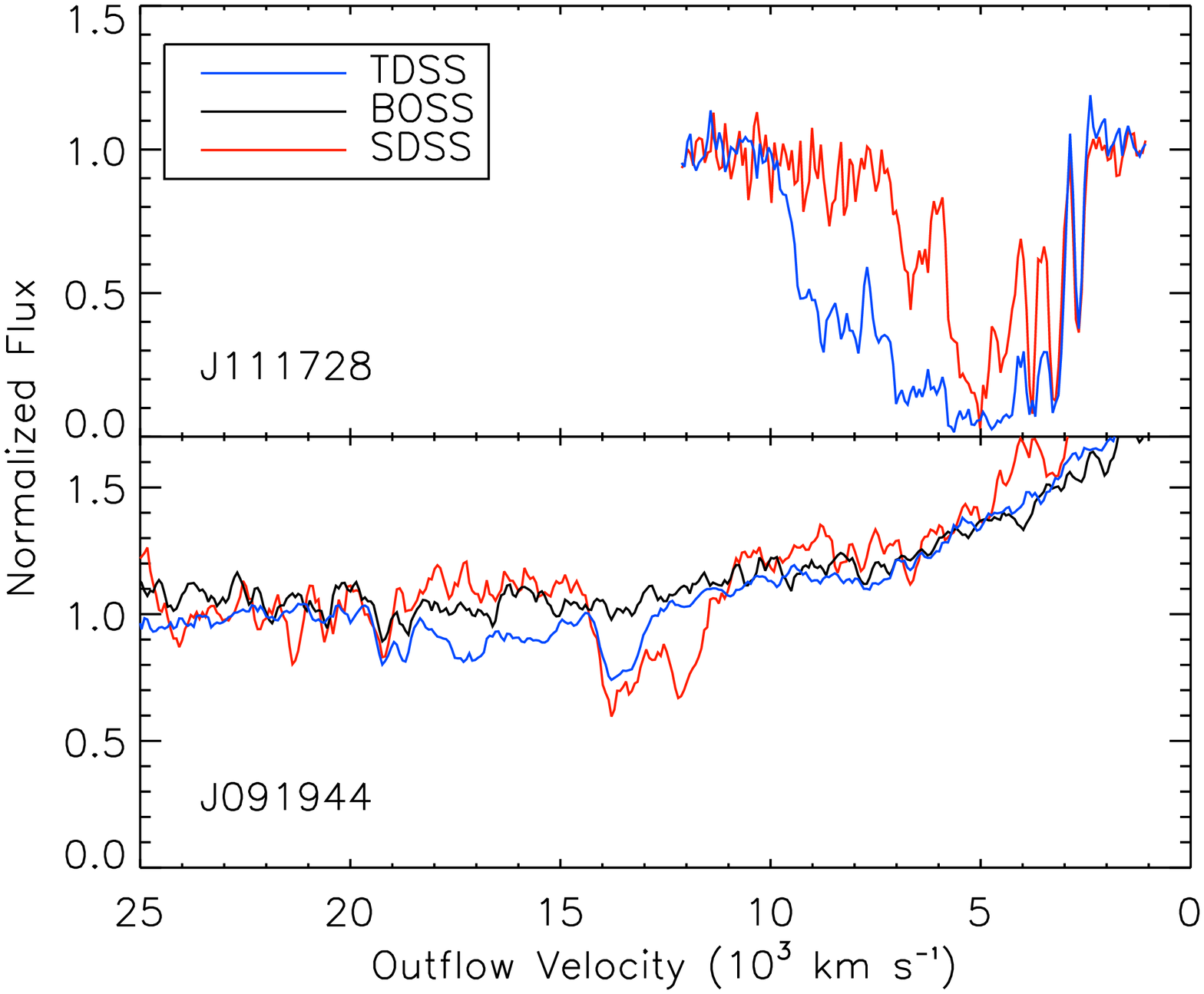}
\plotone{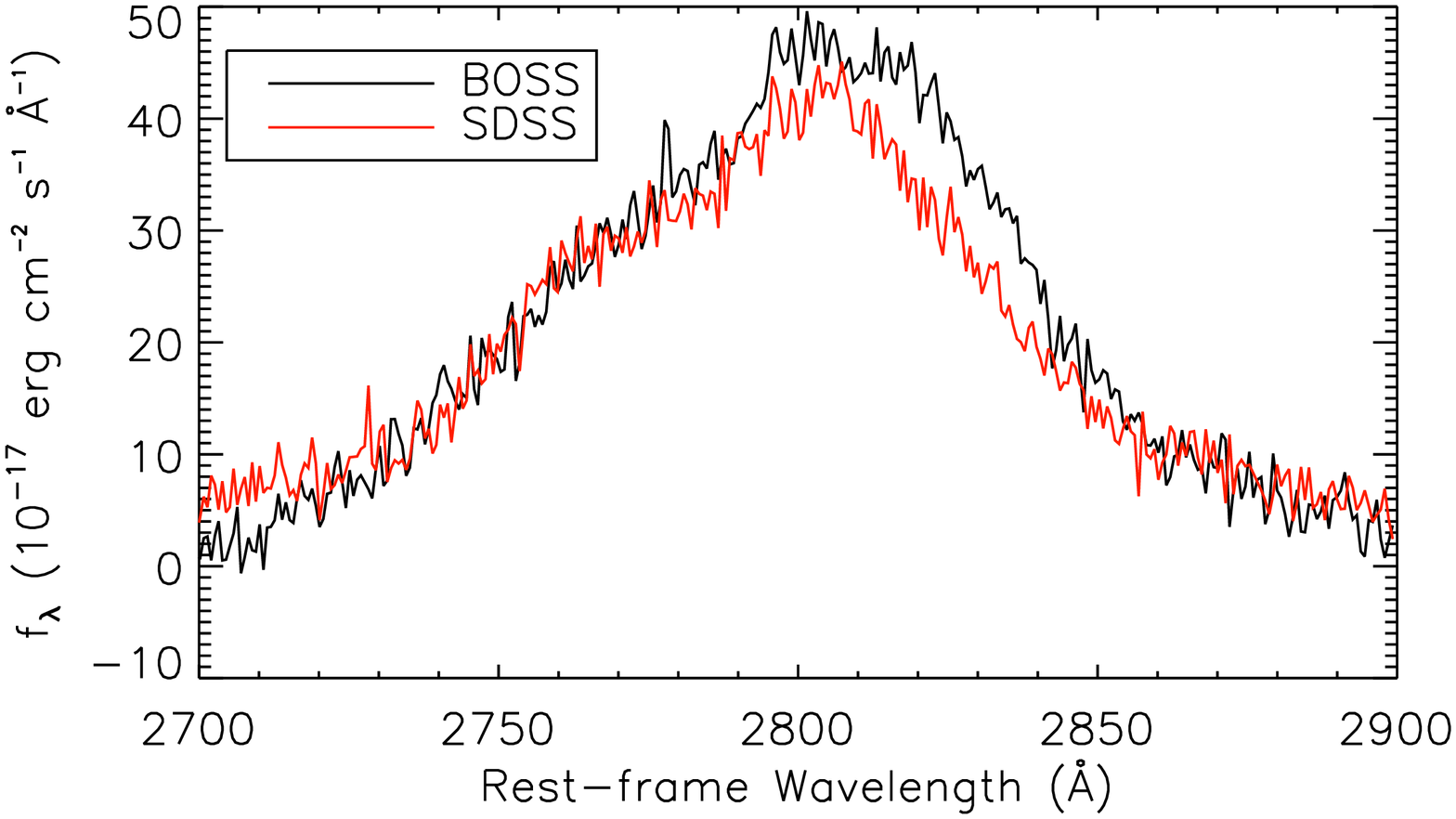}
\caption{{\it Top:} Example spectra of BAL troughs from TDSS (target class {\tt
    TDSS\_FES\_VARBAL}, \S\ref{sec:fes:VARBAL}). 
  The  \ion{C}{4} BAL troughs for quasars SDSS~J111728.75+490216.4 \citep{gri16} and
  SDSS~J091944.53+560243.3 \citep{mcgraw17} are displayed in the top
  and bottom panels respectively, where the velocity is relative to the
  rest frame wavelength of \ion{C}{4}. In the top panel, the
  spectroscopic MJDs are 57129 (TDSS) and 52438 (SDSS); in the
  bottom panel they are 57346 (TDSS), 56625 (BOSS), and 51908 (SDSS).
  The lower panel shows an example of a BAL re-emergence. 
{\it Bottom:} SDSS~J163709.31+414030.8, a candidate SBHB at $z=0.760$
from \cite{wang16} showing a \ion{Mg}{2} velocity shift similar to those in the target class {\tt
    TDSS\_FES\_MGII} (\S\ref{sec:fes:MGII}).
\label{fig:bal}}
\end{figure}

Figure~\ref{fig:bal} shows two examples of a variable BAL quasars
observed in TDSS. 
The selection recipes used to obtain these targets are detailed 
below in a step-by-step manner. 

\subsubsection{Main BAL sample}

The steps to select the majority of FES BAL quasars are as follows:

\begin{enumerate}
\item Match the DR5 BAL catalog \citep{gib09} to the DR5Q. This
  catalog provides full positional, photometric, and spectroscopic 
   information for each BAL quasar.  Positions agree to within
   $0\farcs1$ as expected.\footnote{Two quasars have different
     redshifts between the two catalogs: J100424.88$+$122922.2 and J153029.05$+$553247.9.
   These inconsistencies are explained on page 759, column 2 of \citet{gib09}.}
   For the TDSS targeting, we adopt the astrometry as measured in SDSS
   DR9 \citep{DR9} for these objects.

\item Choose BAL quasars with $i<19.28$. 
   These $i$ magnitudes in the DR5 quasar catalog are not corrected for 
   Galactic extinction, which is generally mild. 

\item From the BAL quasars chosen in step 2, we only accept those with
  ${\rm BI_0}> 100$~km~s$^{-1}$ in one of their BAL troughs. Here,
  ${\rm BI_0}$ is the modified balnicity index defined in \citet{gib09}.
   This cut removes weak BALs that could have
   been mis-classified due to e.g., underlying continuum
   uncertainties. 

   We also constrain redshifts as follows \citep[see Section 4 of][]{gib09}: 
       \emph{(i)}   1.96--5.55 for \ion{Si}{4} BALs;
       \emph{(ii)}  1.68--4.93 for \ion{C}{4}  BALs;
       \emph{(iii)} 1.23--3.93 for \ion{Al}{3} BALs; and
       \emph{(iv)}  0.48--2.28 for \ion{Mg}{2} BALs.
   If a BAL quasar with troughs from multiple ions satisfies any one of these
   required redshift ranges, then it is accepted.

\item For the objects with coverage in the rest-frame window 1650 --
  1750\AA, we only consider those with {\tt SNR\_1700}$ \ge 6$,
  where  {\tt SNR\_1700} is the SNR measurement in this wavelength
  window from the DR5 BAL catalog. This cut  ensures a high-quality 
   first-epoch spectrum for comparison purposes. 
 The resulting number of BAL quasars is 2005.

\item At this point, a manual identification of 476 supplemental BAL targets 
   was performed (led by author P.\ B.\ Hall). These targets may violate one or more 
   of the above selection criteria, but have been identified as worthy of
   additional study nonetheless.\footnote{No explicit magnitude or SNR
     cut was made, but a very low SNR 
   spectrum would have had to be quite interesting to be included.}
 They include the following object classes:

\begin{itemize}
   \item     BAL quasars originally detected in the Large Bright
     Quasar Survey \citep{hewett01} or FIRST Bright Quasar Survey
     \citep{white00}, or otherwise having discovery spectra 
    predating SDSS by up to 10 years or more;

    \item Redshifted-Trough BAL quasars \citep{hall13,zhang17}, a rare class for which competing possible
    explanations make different predictions about trough variability;

    \item Overlapping-Trough BAL quasars with nearly complete absorption below Mg II
    at one epoch but which in several cases \citep{hall11,rafiee16} have already
    shown extreme variability;

    \item BAL quasars observed more than once by SDSS and/or BOSS, and thus
    already possessing more than one epoch for comparison to SDSS-IV,
    including objects with BAL troughs which emerged between SDSS and BOSS;

    \item BAL or X-ray weak quasars selected for their unusual properties where
    observations of future variability (or lack thereof) may
    help determine the processes responsible for their unusual spectra.
\end{itemize}

   After this addition, the resulting number of BAL quasars is 2481
   (2005 regular plus 476 supplemental).

\end{enumerate}

\subsubsection{DR12 Objects}

 To increase the sky coverage of BAL targets, we employed a similar target
 selection as before for the BALs in a preliminary version of DR12Q
 from 22 March 2014 (I.\ P{\^a}ris 2014, private communication). 
   To select BAL targets from this database, we 
   focus only on \ion{C}{4} BAL selection, since this is arguably the primary ion of
     interest (and the one for which we had the needed data
     for selection). We require: 
\begin{enumerate}
   \item   Magnitude $i < 19.8$. 

   \item   ${\rm BI}> 100$~km~s$^{-1}$, where BI is the
     balnicity index defined by the parameter {\tt BI\_CIV} in \citet{par16}.

   \item   BAL visual inspection flag to be positive ({\tt BAL\_FLAG\_VI} $= 1$;
     this cut only dismisses a few objects satisfying the {\tt BI\_CIV} $> 100$~km~s$^{-1}$
     requirement and thus is a small effect).

   \item  A redshift range $z = 1.68$--4.93, which provides complete 
     coverage of the \ion{C}{4} BAL region. 

   \item  The coordinates are within $-50^{\circ}<\alpha<50^{\circ}$ and
     $17.5^{\circ}<\delta<60^{\circ}$ (see above). 
\end{enumerate}
   Application of the criteria above produces 294 targets. 

   Finally, a manual identification of 313 additional special BAL 
   targets was performed (led by author P.\ B.\ Hall). These targets may violate 
   one or more of the above selection criteria, but have been identified 
   as critical for study nonetheless. These objects were selected
   in two different ways:
\begin{enumerate}
    \item 307 quasars were selected from the preliminary DR12 quasar catalog.
     All targets have ({\tt BI\_CIV} $> 0$~km~s$^{-1}$) or ({\tt BAL\_FLAG\_VI} $= 1$) and one or 
     more of the following:
      \emph{(i)} OVI coverage (and preferentially narrow troughs); 
      \emph{(ii)} A high-velocity \ion{C}{4} trough ($> 30,000$ km~s$^{-1}$); 
      \emph{(iii)} Possible redshifted absorption; 
      \emph{(iv)} An existing SDSS spectrum as well as a BOSS
      spectrum; and
      \emph{(v)} Some other unusual property, thus classifying it as
      an ``odd-BAL''.

   \item  Six known quasars were selected from the printed catalogs of 
      \citet{jun91,jun92} or \citet{sow97}.
\end{enumerate}
 In total, thus, there are $294 + 313 = 607$ BAL quasars from this
 second pass of BAL targeting.

\subsection{Candidate Supermassive Binary Black Holes Based on Shifted
  \ion{Mg}{2} Lines {\tt (TDSS\_FES\_MGII)}}
\label{sec:fes:MGII} 

Supermassive black hole binaries (SBHBs) are throught to be a common
consequence of the merger of two massive galaxies.  According to the
evolutionary scenario described by \citet{beg80}, sometime after the merger of
the parent galaxies, the two black holes form a bound binary whose
separation decays first by dynamical friction, then by scattering of
stars, and finally by the emission of gravitational radiation.\footnote{There may be an additional phase before the emission of gravitational waves where the binary separation decays via interactions between the binary and a gaseous 
disk.} The slowest
stage in this evolutionary scheme is thought to correspond to an
orbital separation of $0.01~{\rm pc}\lesssim a\lesssim 1\;$pc. Thus, observational efforts
have focused on finding SBHBs at these orbital separations using
RV variations of the broad emission lines \citep[by
  analogy with double-lined or single-lined spectroscopic binary
  stars; e.g.,][]{gaskell83,gaskell96}.  So far, direct observational
evidence for SBHBs with two active BHs via this method has been
elusive \citep[e.g.,][]{eracleous97,liuEH16}.  Recent surveys have
concentrated on candidates SBHBs with one active black hole and have
utilized the large samples of quasar spectra available in the SDSS
archive \citep{tsalmantza11, eracleous12, shen13, ju13, liu14,
  runnoe17}. The general strategy of these surveys is to select
quasars whose broad Balmer or \ion{Mg}{2} lines are offset from the
frame defined by the narrow lines by $\sim 1000\;{\rm km\;s^{-1}}$ or
more and/or search for systematic RV variations between
the first-epoch spectra and spectra taken several years later.

This program is a continuation of the work of \citet{ju13} who studied
the broad \ion{Mg}{2} emission lines of $0.36<z<2$ quasars with
multiple SDSS observations.  The spectra from this program can be used
to detect velocity shifts in SBHBs with separations of $\sim0.1\;$pc
and orbital periods of $\sim$100 years, assuming that the BHs have
masses of order $10^9\;{\rm M}_{\odot}$.

From the sample of all quasars in DR7Q with
multiple SDSS spectra of the \ion{Mg}{2} line, \citet{ju13} identified
seven robust SBHB candidates along with 57 more candidates that were
less secure, for a total of 64 targets.  The program is designed to
obtain a third-epoch spectrum for all candidates, with highest
priority given to the seven robust candidates, in order to search for
monotonic velocity shifts relative to first epoch. The first results from this
program were reported in \citet{wang16}, in which the authors rule out a binary model for 
the bulk of candidates by comparing the variations in the velocity
shifts over 1-2~yr and 10~yr.  They also find that $\lesssim 1$\%  of
active SMBHs reside in binaries with $\sim$0.1~pc separations  
observed in TDSS.  The example shown in the bottom panel of
Figure~\ref{fig:bal} is a candidate from \citet{wang16} 
with a prominent line shift.

\subsection{Variability of Disk-Like Broad Balmer Lines {\tt (TDSS\_FES\_DE)}} 
\label{sec:fes:DE} 

Broad Balmer lines with double-peaks, twin shoulders, or flat tops can
be found in about 15\% of radio-loud AGNs at $z<0.4$
\citep{eracleous94,eracleous03} and in about 3\% of AGNs at $z<0.33$
in the SDSS \citep{str03}, depending on radio-loudness and possibly Eddington ratio. Although a number of ideas have been discussed in the
literature for the origin of these line profiles, a physical model
attributing the emission to the outer parts of the accretion disk is the
most successful in explaining the Balmer line profiles and other
properties of these objects \citep[see discussion in][and references
  therein]{eracleous94,eracleous03,eracleous09}. Thus, we refer to
these objects as disk-like emitters hereafter. Previous long-term
monitoring of disk-like emitters has sampled about two dozen objects
over 20 years \citep[e.g.,][and references therein]{storchi03,
  gezari07, flohic08, lewis10, popovic11, popovic14,sergeev00,
  sergeev17} at $z<0.4$, most of which are radio loud.

This FES program expands the scope of past monitoring efforts by
re-observing for at least one more epoch a much larger number of
disk-like emitters drawn from the SDSS. This selection method leads to
a much wider variety of objects than those targeted by previous
campaigns, namely more luminous objects, objects with higher Eddington
ratios, and radio-quiet objects. This program also targets 
objects at $z\sim 0.6$, which are even more luminous than
those at $z<0.4$.

Included in the target list are 1251 objects from DR7Q 
 distributed over $\sim$6300 deg$^2$ (i.e., 0.2~deg$^{-2}$).  The
targets comprise ``classic'' disk-like emitters \citep[at $z<0.33$
  taken from][]{str03} and higher-redshift analogs \citep[$z\sim 0.6$;
  from][]{luo13}, as well as additional objects identified by
\citet{shen11}. A total of 220 objects are ``classic'' disk-like emitters \citep[objects
  whose Balmer profiles can easily be modeled by a rotating accretion
  disk, e.g.,][]{eracleous09} while the remaining
objects have very asymmetric Balmer profiles that can plausibly be
attributed to a perturbed disk (for example one with a prominent
spiral) or to a SBHB (see \S\ref{sec:fes:MGII}). The magnitudes of the
targets are $i < 18.9$. The TDSS spectra will cover H$\alpha$ and
H$\beta$ for the $z<0.4$ objects, and H$\beta$ and \ion{Mg}{2} for the
$z\sim 0.6$ objects.  The time baseline will be $> 10$~years for most
objects. The 1251 targets of this program include 28 objects
identified as promising sub-pc binary SMBH candidates with observed
H$\beta$ line shifts between two epochs in SDSS-I/II from
\citet{shen13}. 

By combining existing SDSS spectra and spectra collected during TDSS,
this program aims to address the following scientific goals. 
First, the observations will empirically characterize the variability of the broad emission
  line profiles, i.e., determine what property of the profiles is
  varying (e.g., width, asymmetry, shift, relative strengths and
  velocities of the peaks or shoulders), as well as the magnitude and
  time scale of the variations.
Second, the data will be compared to a wide array of  models of disk
  perturbations, including warps, self-gravitating clumps, and spiral or
  other waves.
Third, this program aims to determine whether the variations represent systematic drifts of
  the line profiles and evaluate whether these changes are consistent with
  RV shifts due to orbital motion in a SBHB.

\begin{figure}
\epsscale{.7}
\plotone{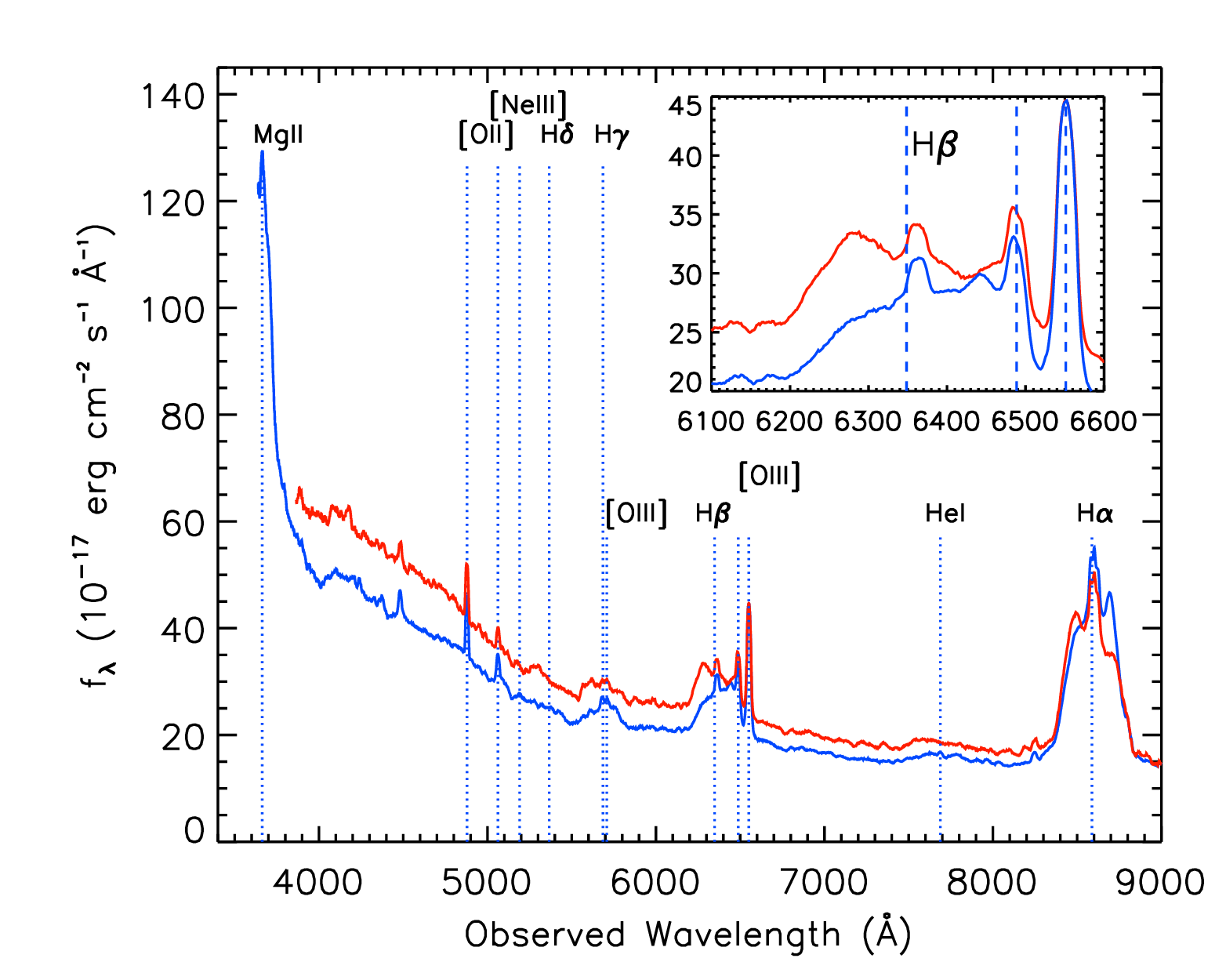}
\plotone{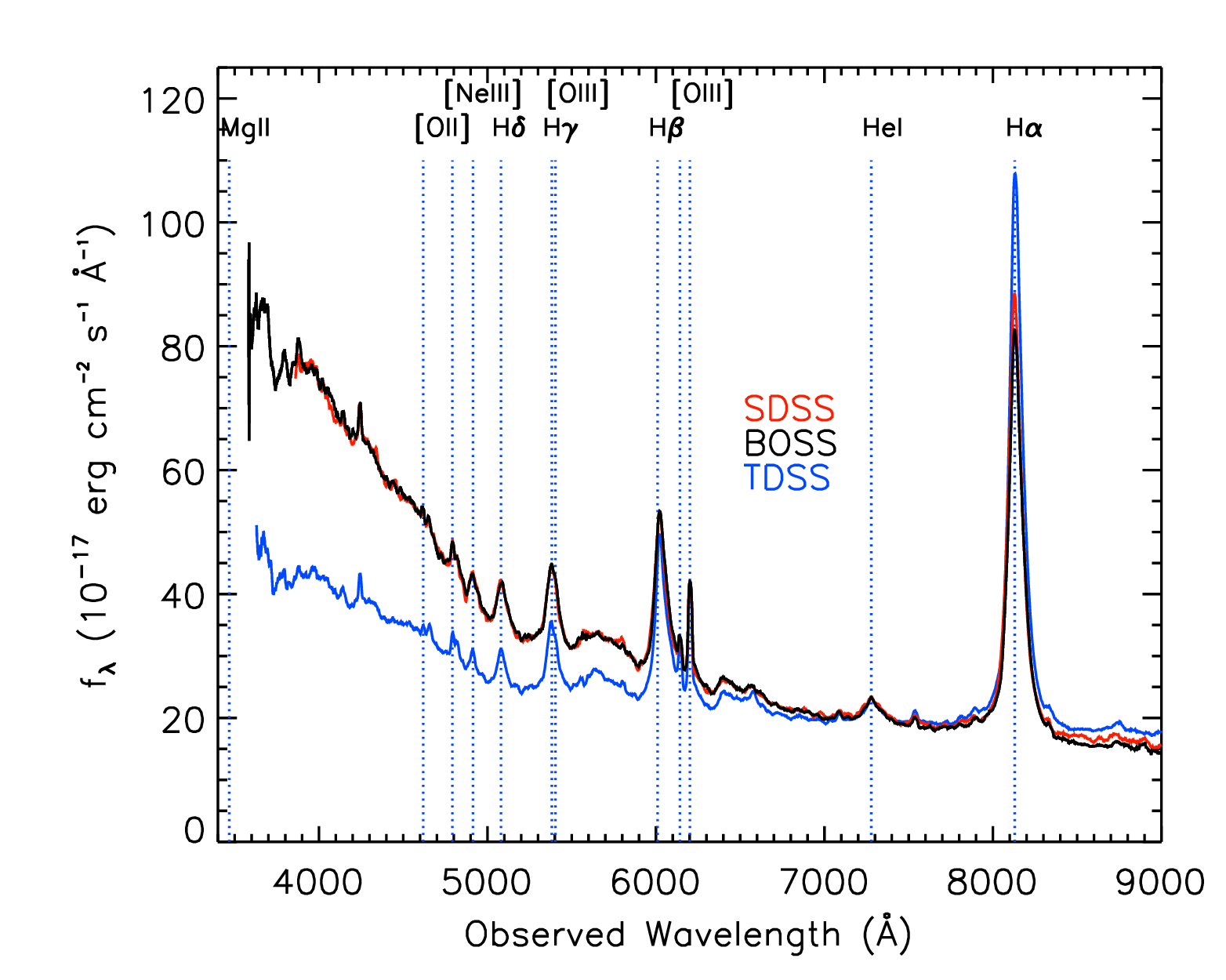}
\caption{{\it Top:} SDSS~J004319.74+005115.4, a disk-like emitter
  quasar at redshift $z=0.308$ (see
  \S\ref{sec:fes:DE}) originally
  observed with the SDSS-I/II spectrograph (red) with an
  additional epoch of spectroscopy from TDSS (blue). The inset shows
  the area surrounding the H$\beta$ emission line. 
  This object has dramatic profile variations over 15~yr (observed
  frame) that may provide clues to the structure and dynamics of the BLR. 
{\it Bottom:} SDSS~J011254.91+000313.0, a $z=0.238$ quasar observed at high SNR  (see
  \S\ref{sec:fes:NQHISN}) that also has a spectrum in BOSS
  (shown in black and very similar to the SDSS spectrum).  The TDSS
  and BOSS spectra (MJDs 57002 and 55214) have been scaled so the 
  flux of [\ion{O}{3}] matches that of the earlier SDSS spectrum (MJD 51794). The  same
  level of smoothing has been applied to SDSS, BOSS, and TDSS spectra,
  and all have an effective wavelength $\lambda_{\rm eff}$=5400\AA.
 \label{fig:DE}}
\end{figure}

An example of disk-like emitter variability seen in one of the targets of this
program is shown in the top panel of Figure~\ref{fig:DE}.

\subsection{Variability of Broad Balmer Lines of Quasars With High-SNR
  Spectra {\tt  (TDSS\_FES\_NQHISN)}}
\label{sec:fes:NQHISN}

This program will yield second (or third) epoch spectra of bright,
low-redshift ($z<0.8$) SDSS quasars with existing high SNR spectra
(requiring that the median SNR per spectral pixel across the full SDSS spectral
range is $>23$). The combination of old and new spectra
will be used to study the general broad line variability of quasars,
including line shape changes and line centroid shifts, on multi-year
timescales. The scientific goals are similar to those of the previous
program (see \S\ref{sec:fes:DE}). In addition to furthering our
understanding of the dynamics of the gas in the broad-line region, the
data from this program will be important for two more applications:
\emph{(i)} a comparison of the variability properties of typical quasar broad
emission lines to the variability properties of disk-like emission
lines (see \S\ref{sec:fes:DE}), and \emph{(ii)} selection of SBHB candidates via
velocity shifts.

The focus of this program is quasars in DR7Q 
 at $z<0.8$. Thus, the spectra will include the H$\beta$
line, as well as the narrow [\ion{O}{3}] doublet that will provide a
reliable redshift and a velocity reference \citep[e.g.,][]{hewett10}. Included in this sample are
1486 quasars with a median SNR~$>23$ per pixel.  

For an example of a quasar targeted in this program, see
Figure~\ref{fig:DE}.    This program is also producing serendipitous discoveries, for example the changing-look quasar from\citet{run16}  was identified from NQHISN spectra.

\section{Repeat Quasar Spectroscopy (RQS)}
\label{sec:rqs}

Quasar variability on multi-year timescales is poorly characterized
for large samples, and our efforts to date have produced unexpected and  
exciting results on the (dis)appearance of broad absorption and emission lines
\citep[e.g.,][]{ak12,run16} as well as large variability of the
continuum and broad line profile shapes. Clearly, in addition to continuing the
existing TDSS programs, a more systematic investigation of quasar
spectroscopic variability is warranted.  As part of the eBOSS ELG
survey \citep{raichoor17}, 
the TDSS was allotted a nominal target density of 10~deg$^{-2}$.   As for 
previous plates, we reserve 10\% of TDSS fibers for the FES programs
described in \S\ref{sec:fes}.   For the remaining fibers, we target {\em
  known} quasars for an additional epoch of spectroscopy  (therefore,  
no SES targets were included on the ELG plates).   
 The target list
includes a magnitude-limited sample of quasars to $i<19.1$, accounting
for the majority of targets ($\gtrsim 7$deg$^{-2}$), and a
variability-selected subsample based on the light curve $\chi^2$, 
favoring quasars with highly-significant photometric variability. 
We also adopt the RQS target selection described here for the eBOSS
plates covering the LRG/quasar targets within Thin82 (chunk20). 

The RQS program is distinct from the SES
TDSS target selection because it \emph{(i)} targets known quasars for repeat
spectroscopy so that spectroscopic variability can be studied, \emph{(ii)}
instead of a pure variability selection, 
includes a complete magnitude-limited sample, since the targets are
already known to be quasars,
\emph{(iii)} uses the full SDSS+PS1 photometric variability information to populate 
 fibers in the S82 region, 
and \emph{(iv)} includes quasars with extended morphologies. 
Quasars with extended morphology are
typically lower-luminosity, lower-redshift sources compared to the
overall SDSS quasar sample, and they have been shown to display
relatively larger variability amplitudes \citep{gal14}. In addition,
by including morphologically extended quasars in RQS, we increase the
redshift/volume overlap with anticipated eROSITA AGN samples \citep{merloni12}. 
The RQS targets also include new SDSS-IV quasars
\citep{myers15,pal16}, which, compared to the previous data 
releases, are on average fainter and extend out to higher redshifts. 

The criteria and priorities $p$ for selecting quasars for repeat
spectroscopy are first described broadly in the enumerated list
below and then in more detail for each sky region.   The FES targets make up the
top-priority TDSS targets ($p=0$) over the entire RQS footprint.  
For the remaining fibers with $p>0$, we identify quasars as follows:

\begin{enumerate}
\item We start with all SDSS quasars drawn from DR7Q and  DR12Q,
  and any new SDSS-IV objects with {\tt CLASS}$=${\tt QSO}.
  Quasars are restricted to  $17<i_{PSF}<21$, where
  $i_{PSF}$ is defined as the median SDSS PSF magnitude.

\item For the majority\footnote{In the NGC region with $126^{\circ}<\alpha\leq 133^{\circ}$, the full list of DR7--12
  quasars, regardless of the number of photometric detections, forms
  our parent sample.} of the selection, our parent sample 
  is quasars with at least two detections in both $g$ and $r$-bands
  among SDSS and PS1 data. To construct
  this parent sample, we consider all primary and secondary
  SDSS photometry within a 1$''$ radius, along
  with PS1 magnitudes measured to better than  $\sigma_{\rm PS1}<0.15$~mag in $g_{\rm PS1}$ and
  $r_{\rm PS1}$, without regard to  morphology or  data quality flags. 

\item Across all sky regions,  the subset with $i_{PSF}<19.1$ defines
  our highest priority RQS targets ($p=1$).
  
\item   Also included in the
  top RQS priority class are $i_{PSF}<20.5$ quasars with multiple existing spectra
  ($N_{spec}>1$), except for the region in Stripe 82 where the density
  for such repeatedly-observed objects exceeds the TDSS fiber density
  allotment.\footnote{The density of $i<20.5$ quasars with $N_{spec}>1$ is 12.75~deg$^{-2}$ and
1~deg$^{-2}$ on and off Stripe 82, respectively.}

\item Within Stripe 82, the SDSS-IV footprint, and a part of the NGC, 
 we use a variability selection to fill the remaining fibers. 
These lower-priority targets are defined by different cuts in
$\chi^2_{pdf}$, the reduced $\chi^2$ for a model for which the quasar's brightness
level does not vary (\S\ref{sec:data:var}).  The same cut is applied in both $g$ and $r$ bands.

\end{enumerate}

Since the density of SDSS quasars varies greatly across the SDSS
footprint, with S82 being the densest, we apply these different cuts
depending on the sky region to achieve an approximately uniform final target
density. The nonuniform coverage of SDSS-IV quasars also alters our
selection method from field to field. After the $i<19.1$  
selection (target flag {\tt TDSS\_RQS1}),  we either use a
variability or magnitude cut to fill the remaining target density depending on the sky region,
where variability-selected targets have a ``v'' appended to the target
flag (e.g., {\tt TDSS\_RQS2v}).\footnote{This choice in target flags
  was made in order to distinguish between second-priority magnitude- and
  variability-selected targets on ELG plates that could potentially include both target
  types.  Note that for the Thin82 plates in eBOSS chunk
20, the variability-selected targets have flag {\tt TDSS\_RQS2}.} A variability cut is especially useful in
regions of high density since a magnitude cut would severly bias the
selection to the brightest sources.  Furthermore, the variability
information is the best in the densest region (S82). Based on the
final target densities, the bulk of the variability selected targets
are in S82.   

The target priorities are enumerated below for each region of the RQS
footprint, where (1) is the highest RQS priority (with target 
flag {\tt TDSS\_RQS1}).  By including objects marked {\tt TDSS\_RQS2}
or {\tt TDSS\_RQS2v}, we 
achieve a rather uniform surface density near the TDSS allotment of
about 10~deg$^{-2}$, although we  supplied targets at a higher density than the nominal
10~deg$^{-2}$ at lower priority to fill in any potential gaps in the ELG target density.\footnote{ 
If ELG targets are dense in a particular region, we might achieve somewhat less than 10~deg$^{-2}$.
For regions where ELG targets are more sparse, TDSS can exceed its nominal 10~deg$^{-2}$ density by using
targets labeled {\tt TDSS\_RQS3} or {\tt TDSS\_RQS3v}, which achieve
densities up to 15~deg$^{-2}$.  We do not supply {\tt TDSS\_RQS3}
targets for the Thin82 plates in eBOSS chunk
20, as all submitted TDSS targets in that area receive a fiber. }
In what follows, we adopt the J2000
coordinates from the DR10 {\tt PhotObj} table:

\begin{itemize}
\item SGC ELG plates: Thin82 (chunk 21: $317^{\circ} < \alpha <
  360^{\circ}$, $-2^{\circ}<\delta < 2^{\circ}$) and Thick82
  (chunk 22: $0 < \alpha < 45^{\circ}$, $-5^{\circ}<\delta < 5^{\circ}$)
\begin{itemize}
\item Region 1 (least dense): off of S82, and currently lacking SDSS-IV
  coverage (see Table~\ref{tab:sdss4})
\\All verified spectroscopic quasars with
\begin{enumerate}
\item $i_{PSF}< 19.1$ or $>1$ existing spectra for $i_{PSF}<20.5$
\item $i_{PSF}< 20.8$, which achieves a surface density near 11~deg$^{-2}$
\item $i_{PSF}< 21  $, which achieves a surface density near 15~deg$^{-2}$
   \end{enumerate}    
\item Region 2: off S82 with SDSS-IV coverage:
\\All verified spectroscopic quasars with
\begin{enumerate}
\item $i_{PSF}< 19.1$ or $>1$ existing spectra for $i_{PSF}<20.5$
\item $\chi^2_{pdf}>27$ for $i_{PSF}<20.5$ to achieve 11~deg$^{-2}$ 
\item $\chi^2_{pdf}>15$ for $i_{PSF}<20.5$ to achieve 15~deg$^{-2}$ 
\end{enumerate}

\item Region 3 (most dense, $|\delta|< 1.3^{\circ}$, in S82): 
\\All verified spectroscopic quasars with
\begin{enumerate}
\item $i_{PSF}< 19.1$
\item $\chi^2_{pdf}>57$ for $i_{PSF}<20.5$ to achieve 11~deg$^{-2}$ 
\item $\chi^2_{pdf}>33$ for $i_{PSF}<20.5$ to achieve 15~deg$^{-2}$ 
\end{enumerate}
\end{itemize}

\item Thin82 (chunk 20: $315^{\circ} < \alpha < 360^{\circ}$, $-2^{\circ} < \delta < 2.75^{\circ}$):

\begin{enumerate}
\item $i_{PSF}< 19.1$
\item $\chi^2_{pdf}>57$, which achieves a surface density near 11~deg$^{-2}$
\end{enumerate}

\item NGC ELG plates (with 376~deg$^2$ already tiled: two rectangles spanning $126^{\circ}<\alpha< 142.5^{\circ}$, $16^{\circ}<\delta < 29^{\circ}$; and $137^{\circ}<\alpha< 157^{\circ}$, $13.8^{\circ}<\delta < 27^{\circ}$)
  \begin{itemize}
  \item  For $133^{\circ} <\alpha< 142.5^{\circ}$, $16^{\circ}<\delta < 29^{\circ}$; and $137^{\circ}<\alpha< 157^{\circ}$, $13.8^{\circ}<\delta < 27^{\circ}$:
    \begin{enumerate}
    \item $i_{PSF}< 19.1$ or $>1$ existing spectra for $i_{PSF}<20.5$:  reaches 10~deg$^{-2}$
    \item $\chi^2_{pdf}>15$ for $i_{PSF}<20.5$ to achieve 12~deg$^{-2}$
    \item $\chi^2_{pdf}>2 $ for $i_{PSF}<20.5$ to achieve 17~deg$^{-2}$
    \end{enumerate}
    
  \item  For all other NGC regions:
    
    \begin{enumerate}
    \item $i_{PSF}< 19.1$:  reaches 10~deg$^{-2}$
    \item $>1$ existing spectra for $i_{PSF}< 20.5$:  reaches 12~deg$^{-2}$
    \item $i_{PSF}< 20$:  reaches 17~deg$^{-2}$
    \end{enumerate}
    
  \end{itemize}
\end{itemize}

All new SDSS-IV objects selected by the above criteria are  visually
confirmed as quasars by inspecting the eBOSS
spectra. We find that this step was mainly necessary for those quasars
with
the {\tt ZWARNING} flag set, but we inspect all selected  SDSS-IV
objects regardless. For objects selected
based on $\chi^2_{pdf}$, we have visually inspected a $3'\times 3'$
SDSS image and rejected objects with close (less than about 5$''$)
neighbors of similar brightness, as well as objects with nearby bright stars
(or extended galaxies) whose diffraction spikes or isophotes
might reasonably contaminate the quasar's photometry.
This image inspection removed about 4-8\%
of candidate targets, depending on the magnitude range.

\begin{figure}
\epsscale{.7}
\plotone{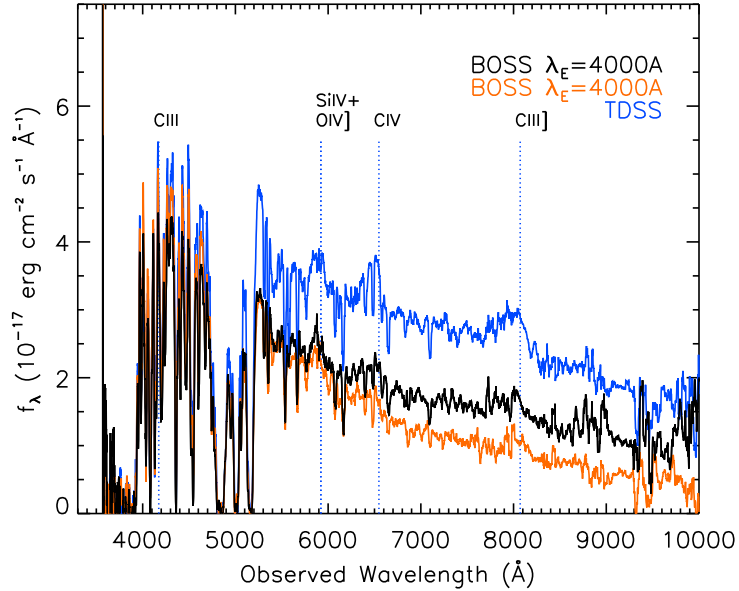}
\caption{ Example spectra of a RQS target observed by TDSS on MJD
  57668 (target class {\tt TDSS\_RQS1}, \S\ref{sec:rqs}).  
  The quasar source has
  four intervening damped Lyman-$\alpha$ systems \citep{sri16}. The
  two previous BOSS spectra of this object are shown, where that shown
  in orange (black) was taken on MJD 55499 (55505). 
\label{fig:rqseg}}
\end{figure}

The approximate anticipated time baselines for RQS targets with
$p=1$--2 are shown in Figure~\ref{fig:timescales}. The recently confirmed
SDSS-IV quasars extend the distribution of probed 
timescales to much shorter baselines. The final distribution of
 all $i<19$, $p=1$ RQS targets ($\approx 11,000$ quasars) in
 rest-frame time baseline-luminosity plane is shown in
 Figure~\ref{fig:dtmi}.  RQS is expected to fill in the large gaps in this plane.
The number of targets in each RQS sky region are listed along with the
numbers for the FES programs in Table~\ref{tab:ntargets}, and an
example RQS target observed in the Fall of 2016 is shown in Figure~\ref{fig:rqseg}.

\floattable
\begin{deluxetable}{llcr}
\tablecaption{Breakdown of FES/RQS Targets \label{tab:ntargets}}
\tablecolumns{6}
\tablewidth{0pt}
\tablehead{
\colhead{Target flag} & \colhead{Description} &
\colhead{No. targets} & \colhead{No. targets} \\
\colhead{} & \colhead{} &
\colhead{submitted} & \colhead{observed} \\
\colhead{} & \colhead{} &
\colhead{} & \colhead{as of DR14} \\
}
\startdata
 TDSS\_RQS1\tablenotemark{a} SGC (chunks 21+22) & Repeat Quasar Spectroscopy on ELG plates & 828+3332 & 0 \\
 TDSS\_RQS2(v)\tablenotemark{b} SGC (chunks 21+22) & Repeat Quasar Spectroscopy on ELG plates & 527+967 & 0 \\
 TDSS\_RQS3(v)\tablenotemark{c} SGC (chunks 21+22) & Repeat Quasar Spectroscopy on ELG plates & 290+667 & 0 \\
 TDSS\_RQS1\tablenotemark{a} NGC (chunk 23)     & Repeat Quasar Spectroscopy on ELG plates & 3366 & 0 \\
 TDSS\_RQS2(v)\tablenotemark{b} NGC (chunk 23)     & Repeat Quasar Spectroscopy on ELG plates & 385  & 0 \\
 TDSS\_RQS3(v)\tablenotemark{c} NGC (chunk 23)     & Repeat Quasar Spectroscopy on ELG plates & 845  & 0 \\
 TDSS\_RQS1\tablenotemark{a} thin82 (chunk 20)  & Repeat Quasar Spectroscopy on LRG/QSO plates & 1064 & 0 \\
 TDSS\_RQS2\tablenotemark{b} thin82 (chunk 20)  & Repeat Quasar Spectroscopy on LRG/QSO plates & 1069 & 0 \\
 TDSS\_FES\_DWARFC   &  Dwarf Carbon stars                       (\S\ref{sec:fes:DWARFC})& 830   & 125 \\         
 TDSS\_FES\_WDDM     &  White dwarf/M dwarf pairs with H$\alpha$ (\S\ref{sec:fes:WDDM})&  402  &  54 \\           
 TDSS\_FES\_ACTSTAR  &  Late-type stars with H$\alpha$           (\S\ref{sec:fes:ACTSTAR})& 1036  & 156 \\
 TDSS\_FES\_HYPSTAR  &  Strongly variable stars                  (\S\ref{sec:fes:HYP})& 1180  & 215 \\                 
 TDSS\_FES\_VARBAL   &  Broad absorption line QSOs               (\S\ref{sec:fes:VARBAL})& 3032  & 950 \\         
 TDSS\_FES\_DE       &  QSO disk-like emitters                   (\S\ref{sec:fes:DE})& 1251  & 232  \\
 TDSS\_FES\_MGII     &  Binary AGN candidates                    (\S\ref{sec:fes:MGII})&   64  & 27 \\                 
 TDSS\_FES\_NQHISN   &  QSOs with previous spectral SNR$>25$     (\S\ref{sec:fes:NQHISN})& 1486  & 324  \\           
 TDSS\_FES\_HYPQSO   &  Strongly variable QSOs                   (\S\ref{sec:fes:HYP})& 1692  & 364  \\                  
\enddata
\tablenotetext{a}{First priority RQS objects ($i<19.1$; see \S\ref{sec:rqs}).}
\tablenotetext{b}{Second priority RQS object (may or may not be
  variability selected).} 
\tablenotetext{c}{Third priority RQS object (may or may not be
  variability selected).}
\tablecomments{Some target groups may overlap. DR14 contains data through
  May 11 2016.  The RQS numbers are restricted to targets already
  assigned to fibers in the 996~deg$^2$ tiled area of the ELG survey so far -- the
  ELG survey is planned to cover  
  an additional 224~deg$^2$ in the NGC by the end of the fourth year
  of SDSS-IV.
All targets are drilled at an effective wavelength of $\lambda_{\rm eff}$=5400\AA. 
  Less than $10$\% of spectra are expected to be of poor quality.
}
\end{deluxetable}

\section{Summary}
\label{sec:summary}

We have described each sample that is targeted for spectroscopic
reobservations in TDSS.   These targets include the FES samples,  
containing stars and quasars of astrophysical interest, and the RQS sample, containing 
a comprehensive set of quasars.  
For the single-epoch ``SES'' TDSS target selection which
constitutes the main TDSS sample, see \citet{mor15}; for initial
results see \citet{rua16}. 
By design, the FES programs total to an average target 
density of 1~deg$^{-2}$ throughout the TDSS survey area (10\%
of all TDSS targets), whereas the RQS targets are prioritized to
achieve a density of $\sim$10~deg$^{-2}$ over a smaller area ($\sim$1200~deg$^{2}$).  
Observations of the SES and FES programs
began in July 2014, and observations of the RQS targets started Fall
2016. TDSS observations are planned to continue until mid-2020, as
part of SDSS-IV eBOSS  \citep[see][]{blanton17}.  By the end of the
survey, TDSS will have obtained reasonable statistical samples for
each FES subclass, containing  at least 100 to 1000 sources each, in
addition to the $\sim13{,}000$ new quasar spectra obtained by the RQS program (see
Table~\ref{tab:ntargets}).

The FES programs span a large range of scientific goals, including:
\begin{itemize}
\item Tracking magnetic field changes over weeks to months
  to  decade-long timescales in M and L dwarfs through observed
  H$\alpha$ variability (\S\ref{sec:fes:ACTSTAR});
\item Investigating the effect of rotation and orbital parameters on stellar
  activity via the H$\alpha$ EW in white dwarf--M dwarf binaries (\S\ref{sec:fes:WDDM});
\item Testing the binary evolution hypothesis for dwarf Carbon stars
  through RV shifts (\S\ref{sec:fes:DWARFC});
\item Exploring the hypervariable star and quasar populations (\S\ref{sec:fes:HYP});
\item Constraining models for BAL variability in quasars through
  analysis of BAL EW and profile variations
over rest-frame timescales of $\approx$4--10~yr, including BAL
acceleration and re-emergence/disappearance events (\S\ref{sec:fes:VARBAL});
\item Searching for velocity shifts in SBHBs with separations of $\sim0.1\;$pc
and orbital periods of $\sim$100 years (\S\ref{sec:fes:MGII});
\item Empirically characterizing the variability of broad emission
  line profiles in disk-like emitting quasars, informing models of disk
  perturbations (\S\ref{sec:fes:DE}); and
\item  Studying the general broad line variability of quasars on multi-year
timescales (\S\ref{sec:fes:NQHISN}).
\end{itemize}

The RQS program, comprising the bulk of the TDSS selection in the ELG
and Thin82 plates, is intended to provide  
spectroscopic variability measurements for an unbiased, high-quality
quasar sample that covers a wide range of redshift and luminosity.    
This dataset will form a complementary sample to the existing quasar sample with repeat
spectroscopy from SDSS and BOSS, which contains 70K quasars as of DR14
(Paris et al.\ 2017), and will span a larger range of timescales. 
Along with sampling long timescales, an important aspect of the FES
and RQS programs is that repeat spectra obtained \emph{within} SDSS-IV, made available
for the most part by the ELG survey \citep{raichoor17}, will sample
short timescales that are also of astrophysical interest.

All the raw and reduced data will be released to the public as part of
the main SDSS-IV data releases. 
The spectroscopic variability studies addressed in these programs will
improve our understanding of the variable sky and facilitate our 
preparation for future sky surveys in the time domain, such as 
Pan-STARRS-2 \citep{burgett12}, the Zwicky Transient Facility
\citep{ztf}, and the Large Synoptic Survey Telescope \citep{lsst}.  
Additionally, the samples presented here will help establish the
targeting strategy for future large-scale spectroscopic variability studies
currently planned for after SDSS-IV.

\acknowledgments
We thank an anonymous referee whose comments greatly improved the manuscript. 
We also thank Lile Wang for provided the data needed for the bottom panel
of Figure 9. Support for this work was provided by the National Aeronautics and
Space Administration through Chandra Award Number AR4-15016X issued by
the Chandra X-ray Observatory Center, which is operated by the
Smithsonian Astrophysical Observatory for and on behalf of the
National Aeronautics Space Administration under contract NAS8-03060.
This material is based upon work supported in part by the National Science Foundation under Grants AST-1715121 and AST-1715763.
WNB, CJG, and SM acknowledge support from NSF grant AST-1516784.
YS acknowledges support from an Alfred P. Sloan Research Fellowship and NSF grant AST-1715579.
Funding for the Sloan Digital Sky Survey IV has been provided by the
Alfred P. Sloan Foundation, the U.S. Department of Energy Office of
Science, and the Participating Institutions. SDSS acknowledges
support and resources from the Center for High-Performance Computing at
the University of Utah. The SDSS web site is www.sdss.org.

SDSS is managed by the Astrophysical Research Consortium for the Participating Institutions of the SDSS Collaboration including the Brazilian Participation Group, the Carnegie Institution for Science, Carnegie Mellon University, the Chilean Participation Group, the French Participation Group, Harvard-Smithsonian Center for Astrophysics, Instituto de Astrof\'{i}sica de Canarias, The Johns Hopkins University, Kavli Institute for the Physics and Mathematics of the Universe (IPMU) / University of Tokyo, Lawrence Berkeley National Laboratory, Leibniz Institut f\"{u}r Astrophysik Potsdam (AIP), Max-Planck-Institut f\"{u}r Astronomie (MPIA Heidelberg), Max-Planck-Institut f\"{u}r Astrophysik (MPA Garching), Max-Planck-Institut f\"{u}r Extraterrestrische Physik (MPE), National Astronomical Observatories of China, New Mexico State University, New York University, University of Notre Dame, Observatório Nacional / MCTI, The Ohio State University, Pennsylvania State University, Shanghai Astronomical Observatory, United Kingdom Participation Group, Universidad Nacional Autónoma de M\'{e}xico, University of Arizona, University of Colorado Boulder, University of Oxford, University of Portsmouth, University of Utah, University of Virginia, University of Washington, University of Wisconsin, Vanderbilt University, and Yale University.

The Pan-STARRS1 Surveys (PS1) and the PS1 public science archive have
been made possible through contributions by the Institute for
Astronomy, the University of Hawaii, the Pan-STARRS Project Office,
the Max-Planck Society and its participating institutes, the Max
Planck Institute for Astronomy, Heidelberg and the Max Planck
Institute for Extraterrestrial Physics, Garching, The Johns Hopkins
University, Durham University, the University of Edinburgh, the
Queen's University Belfast, the Harvard-Smithsonian Center for
Astrophysics, the Las Cumbres Observatory Global Telescope Network
Incorporated, the National Central University of Taiwan, the Space
Telescope Science Institute, the National Aeronautics and Space
Administration under Grant No. NNX08AR22G issued through the Planetary
Science Division of the NASA Science Mission Directorate, the National
Science Foundation Grant No. AST-1238877, the University of Maryland,
Eotvos Lorand University (ELTE), the Los Alamos National Laboratory,
and the Gordon and Betty Moore Foundation.

\vspace{5mm}
\facilities{SDSS, Pan-STARRS}



\appendix
\section{SDSS-IV coverage}

Table~\ref{tab:sdss4} lists the regions of sky relevant to the ELG
survey \citep{raichoor17} that had SDSS-IV coverage at the time of RQS
targeting. This information would be necessary for reproducing the
RQS target selection.

\begin{table}
\centering
\caption{SDSS-IV Coverage as of Summer 2016\label{tab:sdss4}}
\begin{tabular}{cc}
\tablewidth{0pt}
\hline 
\hline 
RA ($^{\circ}$) & Dec ($^{\circ}$)\\
\hline 
\hline
\multicolumn{2}{c}{Boxes spanning ranges:}\\
\hline
$348 <\alpha <357      $& $-1   <\delta < +1  $\\
$\alpha>357            $& $-3.5 <\delta < +7.2$\\
$8.2 <\alpha < 19.4    $& $-8.6 <\delta < -5.9$\\
$0.0 <\alpha \leq 32.8 $& $-5.6 <\delta < +3.1$\\
$32.8<\alpha < 45.0    $& $-6.0 <\delta < +3.1$\\
$0.0 <\alpha <1.75     $& $+4.6 <\delta < +7.0$\\
\hline
\multicolumn{2}{c}{Circles of radius $1.5^{\circ}$ centered on:}\\
\hline
4.3  & +4.0\\
6.8  & +4.0\\
14.5 & +4.0\\
30.0 & +4.0\\
36.7 & +4.0\\
42.3 & +4.0\\
\hline
\end{tabular}
\end{table}





\bibliography{refs} 
\bibliographystyle{aasjournal}




\end{document}